\documentclass[english]{article}
\usepackage{mathptmx}
\usepackage[T1]{fontenc}
\usepackage[latin9]{inputenc}
\usepackage{geometry}
\usepackage{subfigure}
\usepackage{color}
\usepackage{graphicx}

\makeatletter

\providecommand{\tabularnewline}{\\}

\usepackage{axodraw}

\usepackage{babel}
\makeatother

\begin{document}

\title{Down Type Isosinglet Quarks in ATLAS}

\maketitle
\begin{center}
{\large R. Mehdiyev}{\large }%
\footnote{{Universit{\'e} de Montr{\'e}al, D{\'e}partement de Physique, Montr{\'e}al, Canada. }%
}{$^{,}$}%
\footnote{\textcolor{black}{Institute of Physics, Academy of Sciences, Baku, Azerbaijan.}%
}{\large , A. Siodmok}%
\footnote{\textcolor{black}{Jagiellonian University, Physics Department, Cracow, Poland.}%
}{$^{,}$}%
\marginpar{%
}%
\footnote{{CERN, Physics Department, Geneva, Switzerland.}%
}{\large , S. Sultansoy}{$^{2,}$}%
\footnote{{TOBB  University of Economics and Technology, Physics Department, Ankara, Turkey.}%
}{\large , G. Unel}%
\footnote{{CERN, Physics Department, Geneva, Switzerland.}%
}{$^{,}$}%
\footnote{{University of California at Irvine, Physics Department, USA. }%
} 
\par\end{center}

\begin{flushright}
ATL-PHYS-PUB-2007-012 \\
SN-ATLAS-2007-067
\par\end{flushright}

\begin{abstract}
We evaluate the discovery reach of the ATLAS experiment for down type
isosinglet quarks, $D$, using both their neutral and charged decay
channels, namely the process $pp\to D\bar{D}+X$ with subsequent decays
resulting in $2\ell+2j+E\!\!\!\!\!/_{T}$, $3\ell+2j+E\!\!\!\!\!/_{T}$
and $2\ell+4j$ final states. The integrated luminosity required for
observation of a heavy quark is estimated for a mass range between
600 and 1000 GeV using the combination of results from different search
channels.
\end{abstract}

\section{Introduction}

In an effort to unify the forces known to the Standard Model (SM)
within a Grand Unified Theory (GUT), one is led to consider extended
families of fermions. One of the often quoted GUT symmetry groups
having $SU_{C}(3)\times SU_{W}(2)\times U_{Y}(1)$ of the SM as a
subgroup is the exceptional Lie group $E_{6}$ with its 27 dimensional
fundamental representation \cite{R-e6},\cite{R-sugra}. Such a large
symmetry group not only puts the known quarks and leptons in the same
family, but also predicts additional particles, such as iso-singlet
quarks and new heavy gauge bosons. In the literature, the new charge
q=$-1/3$ iso-singlet quarks are denoted by the letters $D$, $S$,
and $B$, representing the additions to the first, second and third
SM families, respectively. 

Here, we investigate the discovery potential of the ATLAS experiment
for the down type isosinglet quark of the first SM family ($D$ quark)
at the forthcoming LHC accelerator. In accordance with the observed
quark mass hierarchy, we will assume the first family's isosinglet
quark $D$ to be the lightest of the particles predicted by the $E_{6}$
group and hence the first one accessible at the LHC energy. It is
assumed that the other new particles predicted by the model, being
very massive, do not contribute to the $D$ quark search. The present
experimental limit on the mass of such an iso-singlet quark is $m_{D}>199$~GeV~\cite{PDG}.
A second assumption will be that the intra-family mixing is stronger
than the mixing between families. Although we will concentrate only
on the $D$ quark, the results will stay valid also for the $S$ quark,
as jets from first and second quark generations are experimentally
indistinguishable. The investigation for the $B$ quark requires detailed
information about the detector response to $b$-jets which is beyond
the scope of this work. 

The Lagrangian relevant for the weak interactions $d$ and $D$ quarks
is given as \cite{R-e6-orhan-metin} :

\begin{eqnarray}
L_{D} & = & \frac{\sqrt{4\pi\alpha_{em}}}{2\sqrt{2}\sin\theta_{W}}\left[\bar{u}^{\theta}\gamma_{\alpha}\left(1-\gamma_{5}\right)d\cos\phi+\bar{u}^{\theta}\gamma_{\alpha}\left(1-\gamma_{5}\right)D\sin\phi\right]W^{\alpha}\label{lagrangian}\\
 & - & \frac{\sqrt{4\pi\alpha_{em}}}{4\sin\theta_{W}}\left[\frac{\sin\phi\cos\phi}{\cos\theta_{W}}\bar{d}\gamma_{\alpha}\left(1-\gamma_{5}\right)D\right]Z^{\alpha}\nonumber \\
 & - & \frac{\sqrt{4\pi\alpha_{em}}}{12\cos\theta_{W}\sin\theta_{W}}\left[\bar{D}\gamma_{\alpha}\left(4\sin^{2}\theta_{W}-3\sin^{2}\phi(1-\gamma_{5})\right)D+\bar{d}\gamma_{\alpha}\left(4\sin^{2}\theta_{W}-3\cos^{2}\phi(1-\gamma_{5})\right)d\right]Z^{\alpha}\;+h.c.\nonumber \end{eqnarray}

where the superscript $\theta$ represents the usual CKM mixings taken
to be in the up sector for simplicity of calculation, $\theta_{W}$
is the weak angle and $\phi$ is the mixing angle between the $d$
and $D$ quarks. This mixing is responsible for the decay of the $D$
quark. The limits on $\phi$ can be obtained from the current precision
measurements for the 3x3 CKM matrix elements assuming that its 3x4
extension has the sum of the squares of the elements of a row equal
to one. The evaluation of the present \cite{PDG} values yield $|\sin\phi|\leq0.045$.
Although the upper limit is used throughout this work, the results
are essentially insensitive to $\sin\phi$ as the pair production
considered here proceeds mostly via gluon exchange. 

The SM fermions acquire their masses through their interactions with
the iso-doublet Higgs field. The Higgs mechanism can also be preserved
in the $E_{6}$ group structure as an effective theory, although other
alternatives such as dynamical symmetry breaking are also proposed
\cite{DSB}. On the other hand, the origin of the mass of the new
quarks ($D,\, S,\, B$) should be due to another mechanism since they
are isosinglets. For this reason and following the literature\cite{R-sugra},
Eq.~(\ref{lagrangian}) does not contain Higgs related terms. However,
if the mass of the Higgs boson is smaller than the isosinglet quark
mass and if the interaction between the Higgs field and $d$ quark
is considered before the spontaneous symmetry breaking but after $d-D$
quark mixing, a new $D$ quark decay channel becomes available: $D\to h\, d$.
The effect of such interaction, depending on the Higgs boson and $D$
quark masses, is to reduce the BR($D\to W\, u$) from 66\% down to
50\% and BR($D\to Z\, d$) from 33\% down to 25\%, giving way to BR($D\to h\, d$)=25\%
(for details see \cite{E6-higgs} and \cite{Rosner}).

\section{Analysis}

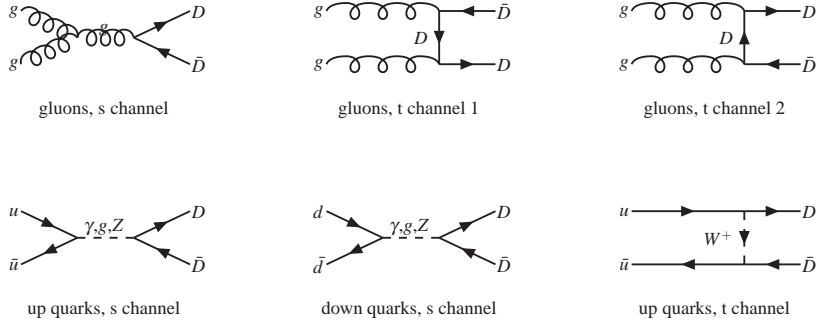
\begin{figure}
\begin{centering}
{
\unitlength=1.0 pt
\SetScale{1.0}
\SetWidth{0.7}      
\scriptsize    
{} \qquad\allowbreak
\begin{picture}(95,79)(0,0)
\Text(15.0,70.0)[r]{$g$}
\Gluon(16.0,70.0)(37.0,60.0){3.0}{3.0} 
\Text(15.0,50.0)[r]{$g$}
\Gluon(16.0,50.0)(37.0,60.0){3.0}{3.0} 
\Text(47.0,61.0)[b]{$g$}
\Gluon(37.0,60.0)(58.0,60.0){3.0}{3.0}
\Text(80.0,70.0)[l]{$D$}
\ArrowLine(58.0,60.0)(79.0,70.0) 
\Text(80.0,50.0)[l]{$\bar{D}$}
\ArrowLine(79.0,50.0)(58.0,60.0) 
\Text(47,30)[b] {gluons, s channel}
\end{picture} \ 
{} \qquad\allowbreak
\begin{picture}(95,79)(0,0)
\Text(15.0,70.0)[r]{$g$}
\Gluon(16.0,70.0)(58.0,70.0){3.0}{3.0} 
\Text(80.0,70.0)[l]{$\bar{D}$}
\ArrowLine(79.0,70.0)(58.0,70.0) 
\Text(54.0,60.0)[r]{$D$}
\ArrowLine(58.0,70.0)(58.0,50.0) 
\Text(15.0,50.0)[r]{$g$}
\Gluon(16.0,50.0)(58.0,50.0){3.0}{3.0} 
\Text(80.0,50.0)[l]{$D$}
\ArrowLine(58.0,50.0)(79.0,50.0) 
\Text(47,30)[b] {gluons, t channel 1}
\end{picture} \ 
{} \qquad\allowbreak
\begin{picture}(95,79)(0,0)
\Text(15.0,70.0)[r]{$g$}
\Gluon(16.0,70.0)(58.0,70.0){3.0}{3.0} 
\Text(80.0,70.0)[l]{$D$}
\ArrowLine(58.0,70.0)(79.0,70.0) 
\Text(54.0,60.0)[r]{$D$}
\ArrowLine(58.0,50.0)(58.0,70.0) 
\Text(15.0,50.0)[r]{$g$}
\Gluon(16.0,50.0)(58.0,50.0){3.0}{3.0} 
\Text(80.0,50.0)[l]{$\bar{D}$}
\ArrowLine(79.0,50.0)(58.0,50.0) 
\Text(47,30)[b] {gluons, t channel 2}
\end{picture} \ 
}
\vskip-1cm
{
\unitlength=1.0 pt
\SetScale{1.0}
\SetWidth{0.7}      
\scriptsize    
{} \qquad\allowbreak
\begin{picture}(95,79)(0,0)
\Text(15.0,70.0)[r]{$u$}
\ArrowLine(16.0,70.0)(37.0,60.0) 
\Text(15.0,50.0)[r]{$\bar{u}$}
\ArrowLine(37.0,60.0)(16.0,50.0) 
\Text(47.0,61.0)[b]{$\gamma$,$g$,$Z$}
\DashLine(37.0,60.0)(58.0,60.0){3.0} 
\Text(80.0,70.0)[l]{$D$}
\ArrowLine(58.0,60.0)(79.0,70.0) 
\Text(80.0,50.0)[l]{$\bar{D}$}
\ArrowLine(79.0,50.0)(58.0,60.0) 
\Text(47,30)[b] {up quarks, s channel}
\end{picture} \ 
{} \qquad\allowbreak
\begin{picture}(95,79)(0,0)
\Text(15.0,70.0)[r]{$d$}
\ArrowLine(16.0,70.0)(37.0,60.0) 
\Text(15.0,50.0)[r]{$\bar{d}$}
\ArrowLine(37.0,60.0)(16.0,50.0) 
\Text(47.0,61.0)[b]{$\gamma$,$g$,$Z$}
\DashLine(37.0,60.0)(58.0,60.0){3.0} 
\Text(80.0,70.0)[l]{$D$}
\ArrowLine(58.0,60.0)(79.0,70.0) 
\Text(80.0,50.0)[l]{$\bar{D}$}
\ArrowLine(79.0,50.0)(58.0,60.0) 
\Text(47,30)[b] {down quarks, s channel}
\end{picture} \ 
{} \qquad\allowbreak
\begin{picture}(95,79)(0,0)
\Text(15.0,70.0)[r]{$u$}
\ArrowLine(16.0,70.0)(58.0,70.0) 
\Text(80.0,70.0)[l]{$D$}
\ArrowLine(58.0,70.0)(79.0,70.0) 
\Text(54.0,60.0)[r]{$W^+$}
\DashArrowLine(58.0,70.0)(58.0,50.0){3.0} 
\Text(15.0,50.0)[r]{$\bar{u}$}
\ArrowLine(58.0,50.0)(16.0,50.0) 
\Text(80.0,50.0)[l]{$\bar{D}$}
\ArrowLine(79.0,50.0)(58.0,50.0) 
\Text(47,30)[b] {up quarks, t channel}
\end{picture} \ 
}
\par\end{centering}

\begin{centering}
\vskip-1cm
\par\end{centering}

\caption{Main diagrams for the pair production of the $D$ quark at the LHC.\label{fig:The-main-diagrams}}

\end{figure}

\begin{figure}
\begin{centering}
\includegraphics[bb=20bp 20bp 555bp 555bp,clip,scale=0.4]{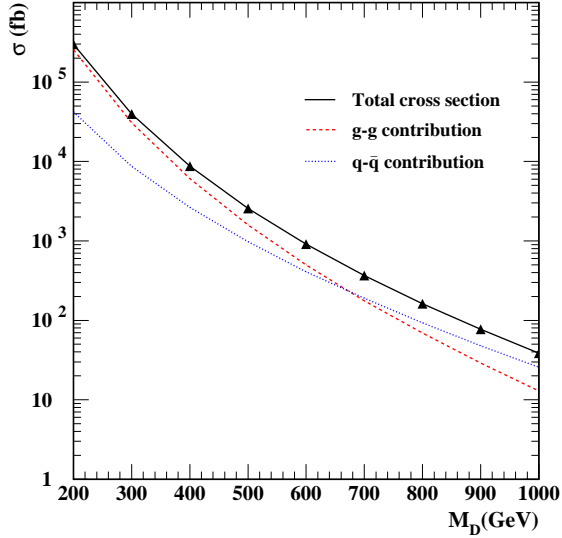}
\par\end{centering}

\caption{Tree level pair production of $D$ quarks at LHC as a function of
the new quark's mass. \label{fig:D-Pair-prod}}

\end{figure}

The leading Feynman diagrams for the pair production of the $D$ quark
at the LHC are given in Fig.~\ref{fig:The-main-diagrams}. The Lagrangian
in Eq. (\ref{lagrangian}) was implemented in two tree level generators:
CompHep \cite{R-calchep} and MadGraph \cite{madgraph}. The pair
production signal cross section was calculated in both generators
as a function of $D$ quark mass, yielding practically equal cross
sections for the same generator level cuts. The result from CompHep
is plotted as a function of the new quark mass in Fig.~\ref{fig:D-Pair-prod}.
Contributions from the electroweak interactions to the production
cross section are found to be much smaller than those from the strong
interactions. Below a $D$ quark mass of about 700 GeV, the dominant
contribution to the production cross section is from the diagrams
in the top row of Fig.~\ref{fig:D-Pair-prod} but as the mass increases,
the quark fusion contribution takes over according to CTEQ6L1 PDFs
\cite{R-cteq}. One should note that single production of $D$ quarks
(in association with a light SM quark) is also possible in this model.
However, since this type of subprocesses, in comparison to the pair
production, would give a different final state the production of single
$D$ quarks is considered in a separate study \cite{E6-single}.

The $E_{6}$ GUT model does not predict the masses of the isosinglet
quarks. Therefore, this study scans some plausible values for the
D quark mass (600 - 1000 GeV) to estimate the experimental reach of
the ATLAS detector \cite{R-atlas-tdr}. With such a large mass, the
isosinglet quarks are expected to immediately decay into SM particles.
A number of decay channels of the $D$ quark pairs are summarized
in Table \ref{tab:Possible-signal-decay-channels}. The process in
the first row has been previously studied and found to be feasible
for discovery up to a mass of about 1 TeV, for an integrated luminosity
of 200 fb$^{-1}$ \cite{NC1-pub}. The present work aims to evaluate
the possibility of discovery of the $D$ quark in the production channels
shown in the remaining rows of the same table using the final states
in column 3. The high transverse momentum of the jets coming from
the $D$ quark decays can be used to identify the signal events. As
for the backgrounds, all the SM tree level processes yielding two
jets and two vector bosons ($Z-Z$ and $W-Z$) are considered. In
this feasibility study, the detector related background events originating
from jet combinatorics and misidentification such as two vector bosons
plus one or three jets are not taken into account. The background
cross section was calculated only in MadGraph as this generator is
suitable to be used on large computing farms. In the numerical calculations,
we set the QCD scale of the signal to the mass of the new quark while
keeping it as the mass of the $Z$ boson for the background, yielding
more conservative results compared to the investigation in \cite{NC1-pub}.
For example, the 3$\sigma$ observation limit gets reduced by 15\%
in the in the $4\ell+2j$ channel studied in that work. The simple
requirements imposed at the generator level are:\begin{eqnarray}
|\eta_{p}| & < & 2.5\;,\label{eq:generator level cuts}\\
p_{T,p} & > & 10\;\mbox{{GeV}}\;,\nonumber \\
R_{p\overline{p}} & > & 0.4\;,\nonumber \\
|\eta_{Z,W}| & < & 5.0\nonumber \end{eqnarray}
 where $R$ is the cone separation angle between two leading partons
($p$) giving rise to two jets. Selection of the pseudorapidity region
is driven by partonic spectra distributions, which are peaked in the
central region. Decays of $Z$ and $W$ bosons were performed by PYTHIA
\cite{R-pythia} which handled the initial and final state radiation
together with the hadronization. The detector response was estimated
using the ATLAS fast simulation program ATLFAST\cite{R-AtlFast} within
the offline analysis framework, ATHENA v11.0.41 . 

\begin{table}
\caption{List of the studied signal channels: the last column contains the
total branching ratios to the final state particles where $\ell$
represents $\mu$ or $e$ .\label{tab:Possible-signal-decay-channels}}

\begin{centering}
\begin{tabular}{c|c|c|c|c}
{\footnotesize $D\bar{{D}}\rightarrow$} & Final State & Expected Signal & Decay B.R. & Total B.R.\tabularnewline
\hline
\hline 
{\footnotesize $Z\, Z\, d\,\bar{{d}}$} & {\footnotesize $Z\rightarrow\ell\bar{{\ell}}$ $Z\rightarrow\ell\bar{{\ell}}$ } & {\footnotesize $4\:\ell\:+2\: jet$} & 0.07$\times$0.07 & 0.0005\tabularnewline
\cline{2-2} \cline{3-3} \cline{4-4} \cline{5-5} 
\multicolumn{1}{c|}{0.33$\times$0.33} & {\footnotesize $Z\rightarrow\ell\bar{{\ell}}$ $Z\rightarrow\nu\,{\bar{\nu}}$ } & {\footnotesize $2\:\ell\:+2\: jet+E\!\!\!/_{T}$} & 2$\times$0.07$\times$ 0.2 & 0.0028\tabularnewline
\cline{2-2} \cline{3-3} \cline{4-4} \cline{5-5} 
 & {\footnotesize $Z\rightarrow\ell\bar{{\ell}}$ $Z\rightarrow q\bar{{q}}$ } & {\footnotesize $2\:\ell\:+4\: jet$} & 2$\times$0.07$\times$0.7 & 0.0107\tabularnewline
\hline 
{\footnotesize $Z\, W\, d\, u$} & {\footnotesize $Z\rightarrow\ell\bar{{\ell}}$ $W\rightarrow l\bar{{\nu}}$ } & {\footnotesize $3\:\ell\:+2\: jet+E\!\!\!/_{T}$} & 0.07$\times$0.21 & 0.0065\tabularnewline
\cline{2-2} \cline{3-3} \cline{4-4} \cline{5-5} 
\multicolumn{1}{c|}{2$\times$0.66$\times$0.33} & {\footnotesize $Z\rightarrow\ell\bar{{\ell}}$ $W\rightarrow q\bar{{q}}$ } & {\footnotesize $2\:\ell\:+4\: jet$} & 0.07$\times$0.68 & 0.0211\tabularnewline
\end{tabular}
\par\end{centering}
\end{table}

\subsection{Search using the $2\ell+2j+E\!\!\!\!\!/_{T}$ channel}

This search aims to exploit the relatively large branching ratio seen
in second line of Table \ref{tab:Possible-signal-decay-channels}.
However the invisible decay of one $Z$ boson, makes reconstruction
of both $D$ quarks impossible. The expected viable signature is two
isolated leptons, two energetic jets and large missing transverse
energy. The reconstructed invariant mass from the two leptons is required
to yield the $Z$ boson mass with an error of $\pm$20~GeV which
is an order of magnitude larger than the reconstruction resolution
$\pm2.5$ GeV. The two most energetic jets are selected as candidates
for the parton daughters of the $D$ quark decays. However, the ambiguity
in the association of a jet and a $Z$ boson, makes each event contribute
twice to the $D$ quark invariant mass histogram, one with the correct
association and one with the wrong one. The background for these events
is estimated by taking into account all the SM processes which give
two energetic jets and two $Z$ bosons, allowing one of them to decay
invisibly and the `other one into electrons or muons. The total cross
section of such processes with the generator level cuts in Eq. (\ref{eq:generator level cuts})
except $p_{T,\, p}>100$ GeV is 2.65 pb. The other SM processes with
$2\ell+2j+E\!\!\!\!\!/_{T}$ background are expected to be negligible,
given the high $E\!\!\!\!\!/_{T}$ in the topology of the signal events,
and the $Z$ boson mass reconstruction requirement. 

The simple cuts for this channel are presented in the set below:

\begin{eqnarray}
\hbox{\#}\,(e,\mu) & = & 2\label{eq:jim-cuts}\\
E\!\!\!/_{T} & \geq & 150\,\mbox{{GeV}}\nonumber \\
M_{\ell\,\bar{\ell}} & = & 90\pm20\:\mbox{{GeV}}\nonumber \\
\#jets & \geq & 2\nonumber \\
p_{T}jet & \geq & 150\,\mbox{{GeV}}\nonumber \end{eqnarray}

\begin{figure}
\begin{centering}
\includegraphics[bb=18bp 10bp 530bp 350bp,clip,scale=0.3]{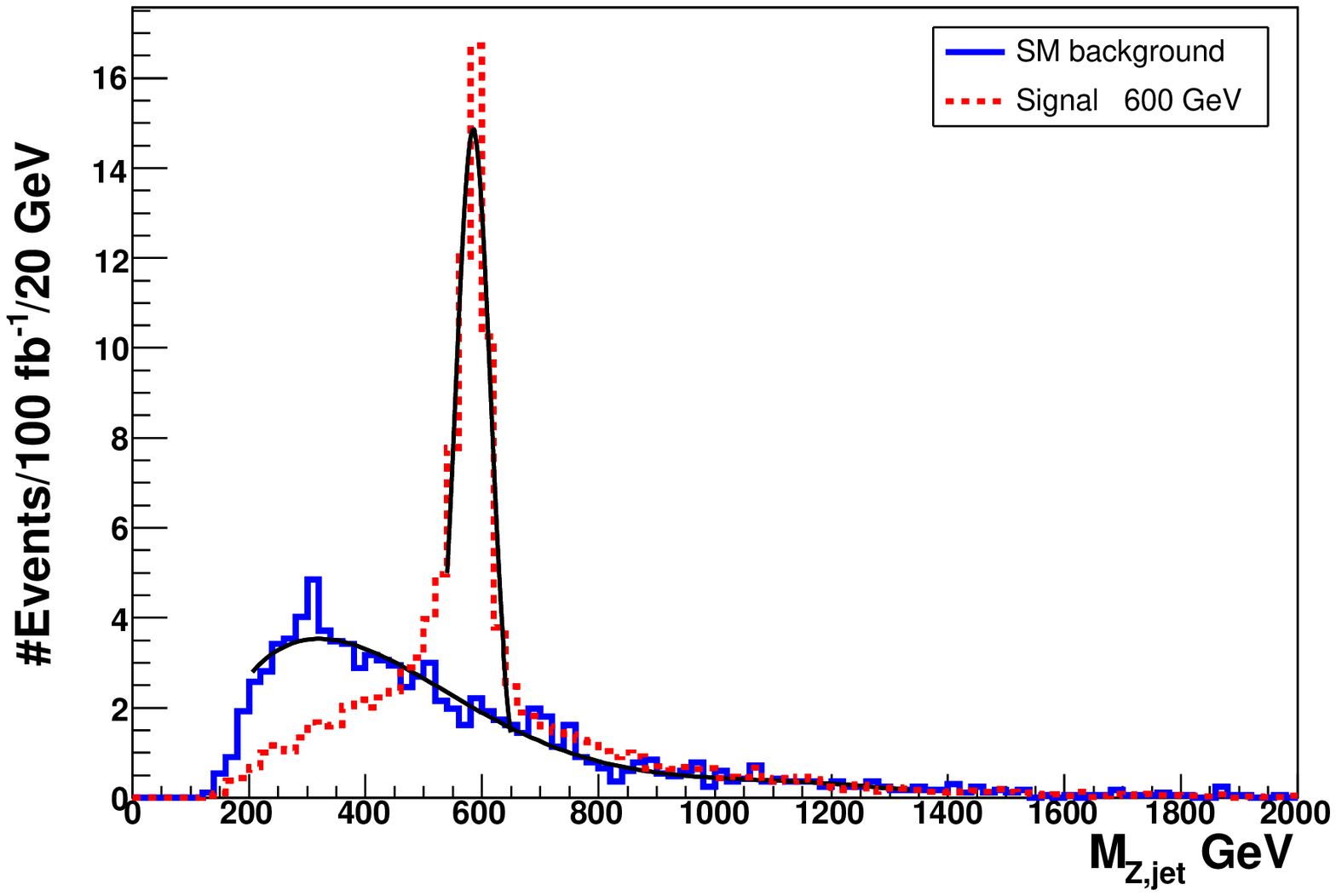}\includegraphics[bb=18bp 10bp 530bp 350bp,clip,scale=0.3]{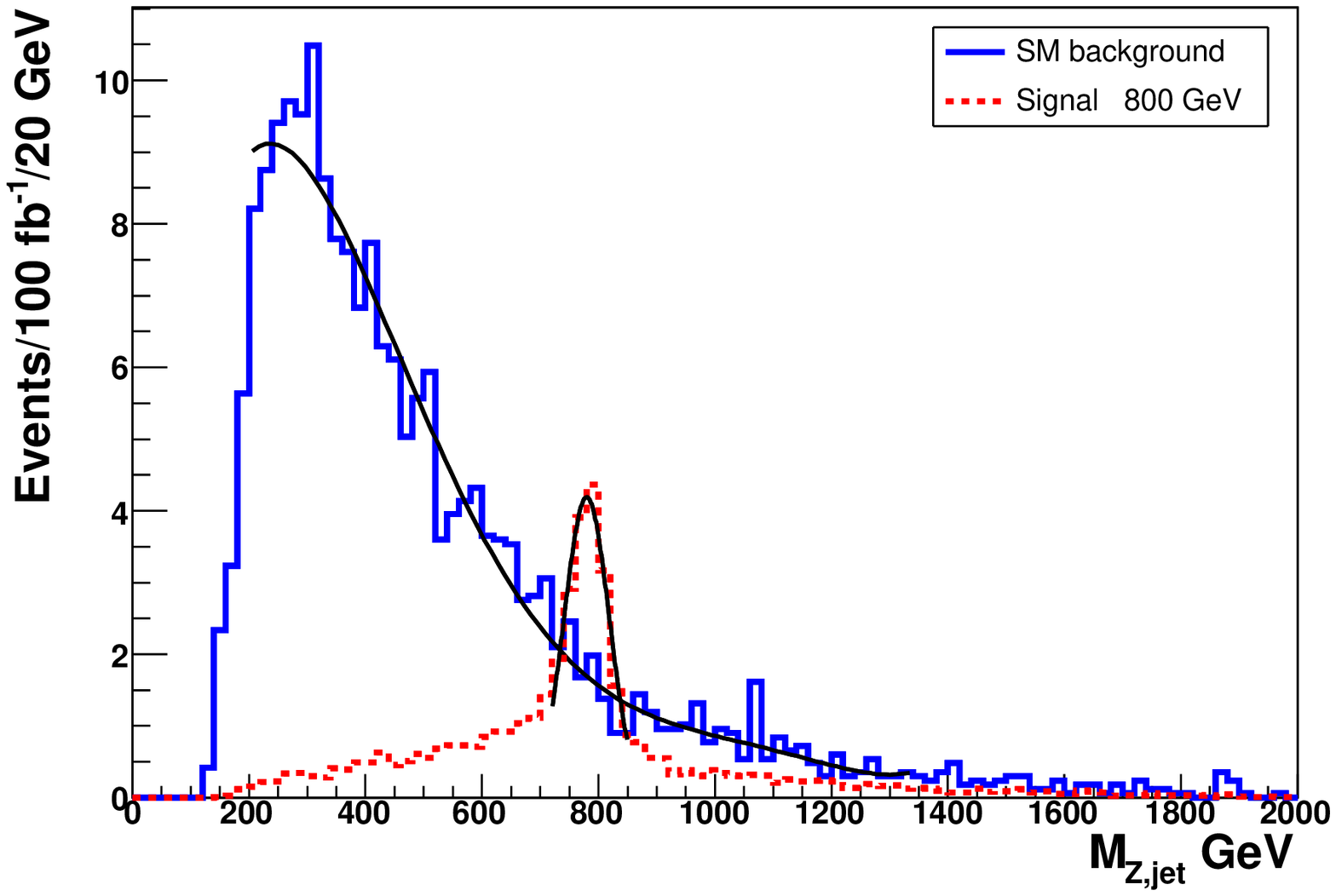}\includegraphics[bb=18bp 10bp 530bp 350bp,clip,scale=0.3]{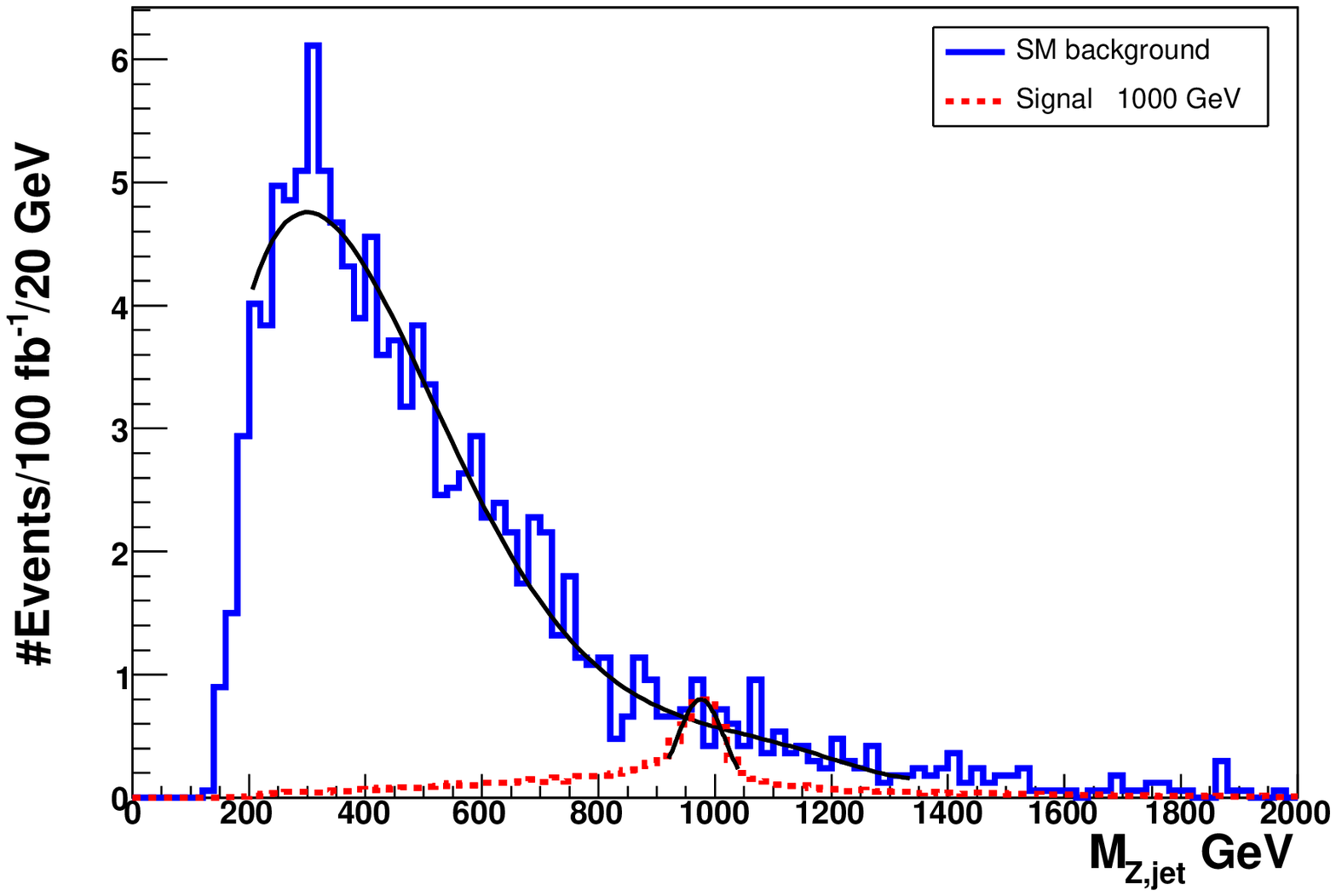}
\par\end{centering}

\caption{Invariant mass spectra of signal and backgrounds for different values
of $D$ quark mass in $2\ell+2j+E\!\!\!/_{T}$ channel\label{fig:Invariant-Mass-NC2}}

\end{figure}

The resulting invariant mass plots for different $D$ quark mass values
are given in Fig.~\ref{fig:Invariant-Mass-NC2} for one year of LHC
data taking which corresponds to an integrated luminosity of 100 fb$^{-1}$.
The solid histograms represent the events originating from the SM
background and the dashed ones from the signal. Both distributions
were fitted separately to reduce statistical fluctuations and to better
estimate the total number of events in the histograms. The number
of events for both signal and background were computed by integrating
the fit functions in the range of $\pm$2$\sigma$ from the center
of the Gaussian fit to the signal peak. Table \ref{tab:signif-NC2}
contains the number of signal ($S$) and background ($B$) events
together with the signal significance calculated in this way for an
integrated luminosity of 100 fb$^{-1}$ . In this note, the signal
significance is calculated as $S/\sqrt{B}$ if $S+B>25$ , otherwise
Poisson statistics are used to obtain the probability P, of compatibility
with the background. In the latter case the significance of the Gaussian
distribution yielding the same probability, P, is given.  

\begin{table*}
\caption{The expected number of events in the mass window around the peak and
the statistical significance for $2\ell+2j+E\!\!\!/_{T}$ channel
after the event selection cuts in equation set \ref{eq:jim-cuts}for
an integrated luminosity of 100 fb$^{-1}$ \label{tab:signif-NC2}.}

\begin{centering}
\begin{tabular}{c|c|c|c}
Mass (GeV) & 600 & 800 & 1000\tabularnewline
\hline
\hline 
S & 53 & 19 & 4\tabularnewline
\hline
B & 12 & 13 & 5\tabularnewline
\hline 
significance & 15.3 & 5.3 & 2.1\tabularnewline
\end{tabular}
\par\end{centering}
\end{table*}

\subsection{Search using the $3\ell+2j+E\!\!\!\!\!/_{T}$ channel}

This channel has one $D$ quark decaying through a $Z$ boson and
another $D$ quark decaying via a $W$ boson, i.e. $DD\rightarrow ZWjj$
final state. The $Z$ boson is expected to be entirely reconstructed
from two electrons or two muons. The $W$ boson can also be partially
reconstructed if a good measurement of missing $E_{T}$ can be achieved.
The signal is sought in the final state made of a pair of energetic
leptons (electrons or muons), accompanied by a third high $p_{T}$
lepton (muon or electron), two high $p_{T}$ jets and the missing
transverse energy.

For the background studies, events consisting of the SM $WZ+2j$ processes
were generated. The background cross section before event selection
is 2.30 pb from the $W^{-}$ case and 3.89 pb from $W^{+}$ case,
summing up to 6.19 pb with the generator level cuts in Eq. (\ref{eq:generator level cuts})
except $p_{T,\, p}>50$ GeV. The composition of background final state
and allowed decay modes of $Z$ and $W$ bosons were chosen to be
exactly the same as for the signal events. The other possible background
contributions due to misidentified or undetected leptons, such as
$ZZ+2j$ or $WW+2j$ events, will be significantly suppressed due
to the event selection: rather high transverse momentum cuts on jets
and the requirement of reconstructing both $Z$ and $W$ bosons in
the same event.

Two cases were separately considered during the event selection for
the lepton composition of the final state. These two cases and the
event selection cuts for a $D$ quark mass of 600 GeV are:

\begin{description}
\item [{A.}] the $Z$ and $W$ decays involving leptons of the same generation
(electrons or muons) were selected using
\end{description}
\begin{eqnarray}
\#e=3 & \hbox{or} & \#\mu=3\label{eq:typeA}\end{eqnarray}

\begin{description}
\item [{B.}] the $Z$ and $W$ decays involving leptons of different generation
(electrons or muons) were selected using
\end{description}
\begin{eqnarray}
\#e=2\;\#\mu=1 & \hbox{or} & \#e=1\;\#\mu=2\,.\label{eq:typeB}\end{eqnarray}

The second case aims to ease the identification of leptons originating
from $Z$ and $W$ bosons. The other analysis cuts are common to both
cases:

\begin{eqnarray}
p_{T,j} & > & 80\:\mbox{{GeV}}\label{eq:cuts_3l_2j}\\
p_{T,e,\mu} & > & 20\:\mbox{{GeV}}\,,\nonumber \\
|\eta(j,\, e\;\hbox{or}\;\mu)| & < & 2.5\,,\nonumber \\
p_{T}^{miss} & > & 30\:\mbox{{GeV}}\,,\nonumber \\
M_{\ell\ell}^{\hbox{rec}} & = & 90\pm20\:\mbox{{GeV}}\,,\nonumber \\
M_{\ell\nu}^{\hbox{rec}} & = & 80\pm20\:\mbox{{GeV}}\,.\nonumber \end{eqnarray}

For higher $D$ quark mass values, the jet transverse momentum cuts
were slightly increased to obtain a favorable significance value wherever
it was statistically possible in both cases.

\begin{figure}
\begin{centering}
\includegraphics[scale=0.3]{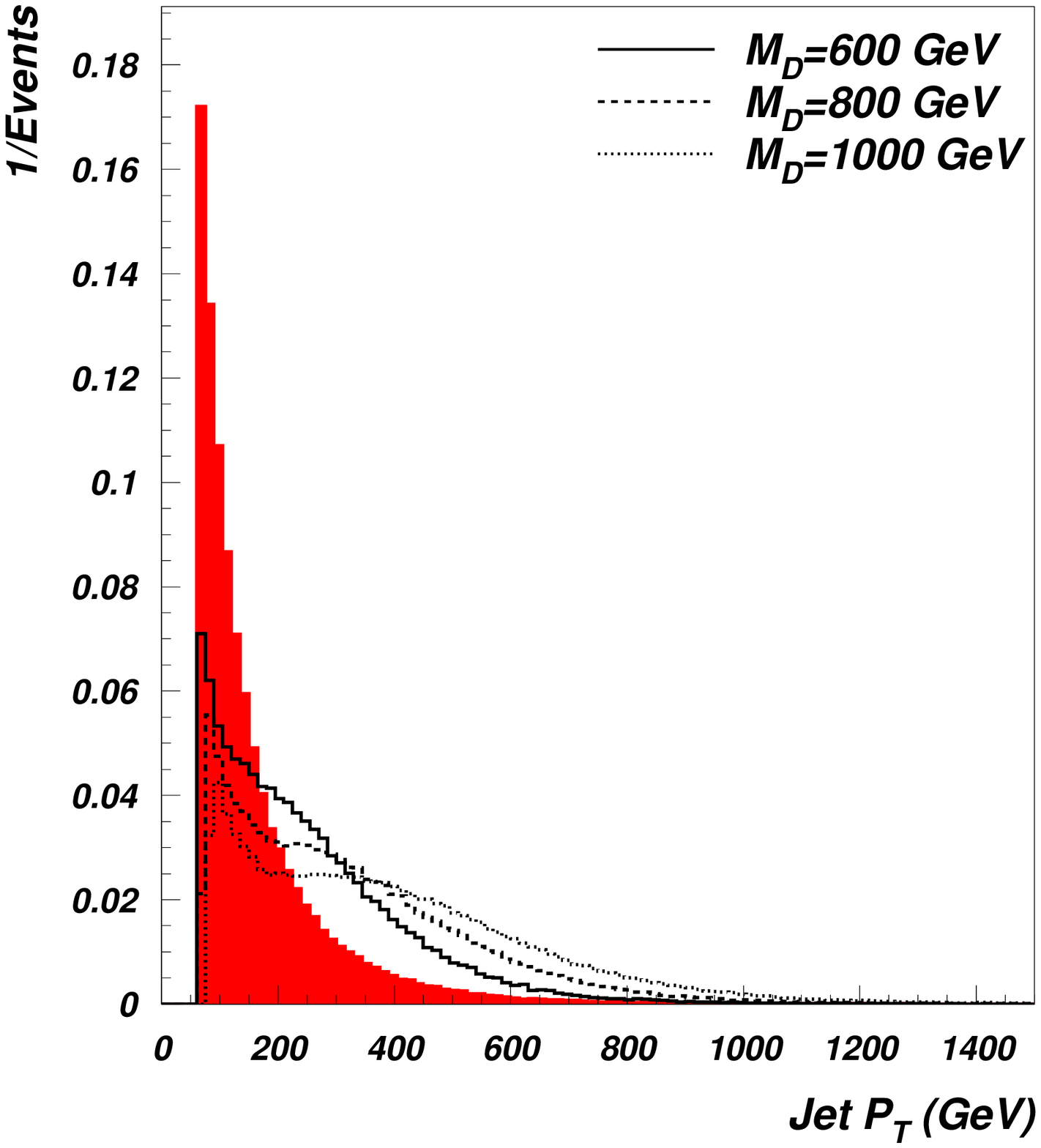}\includegraphics[scale=0.3]{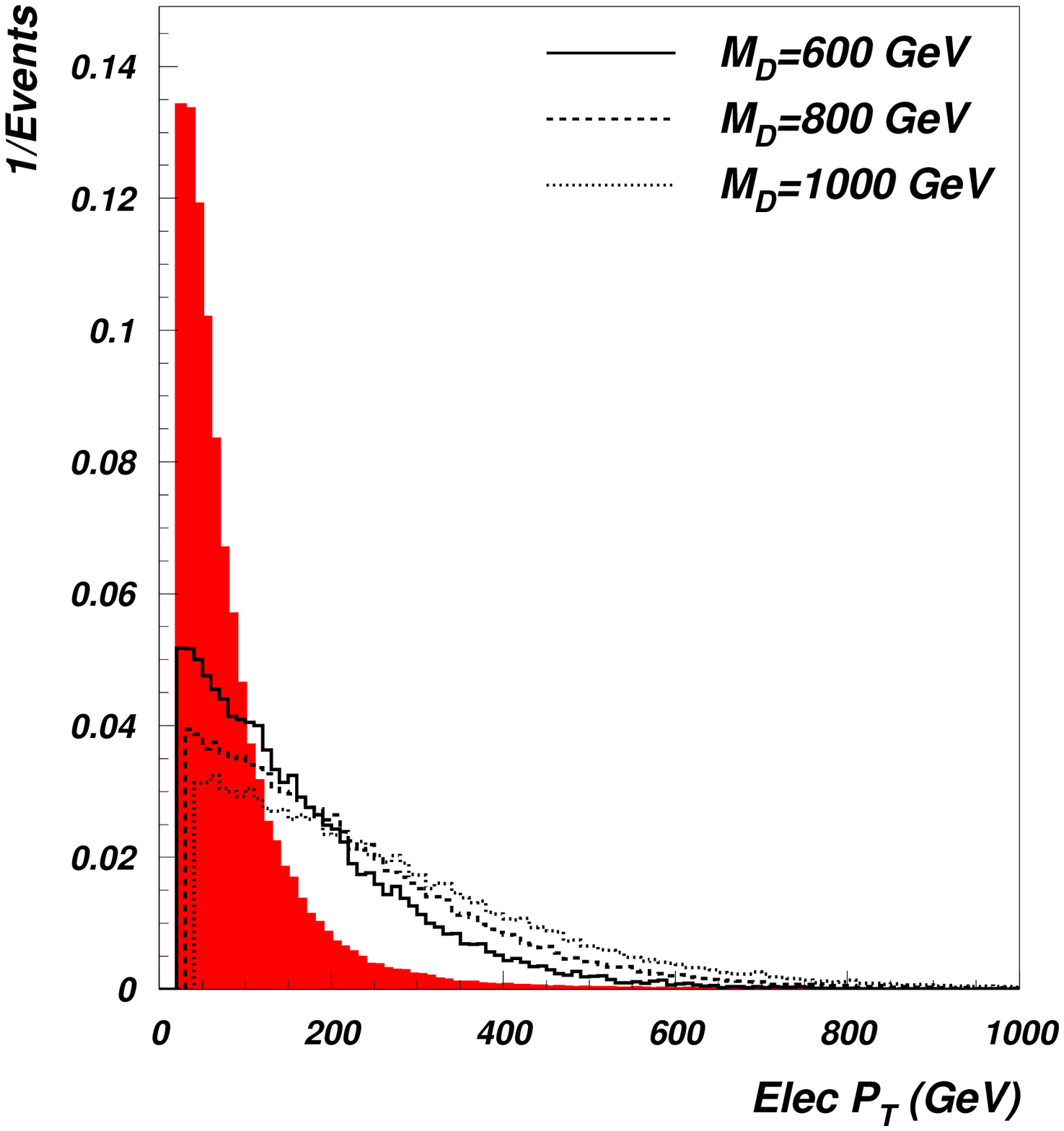}\includegraphics[scale=0.3]{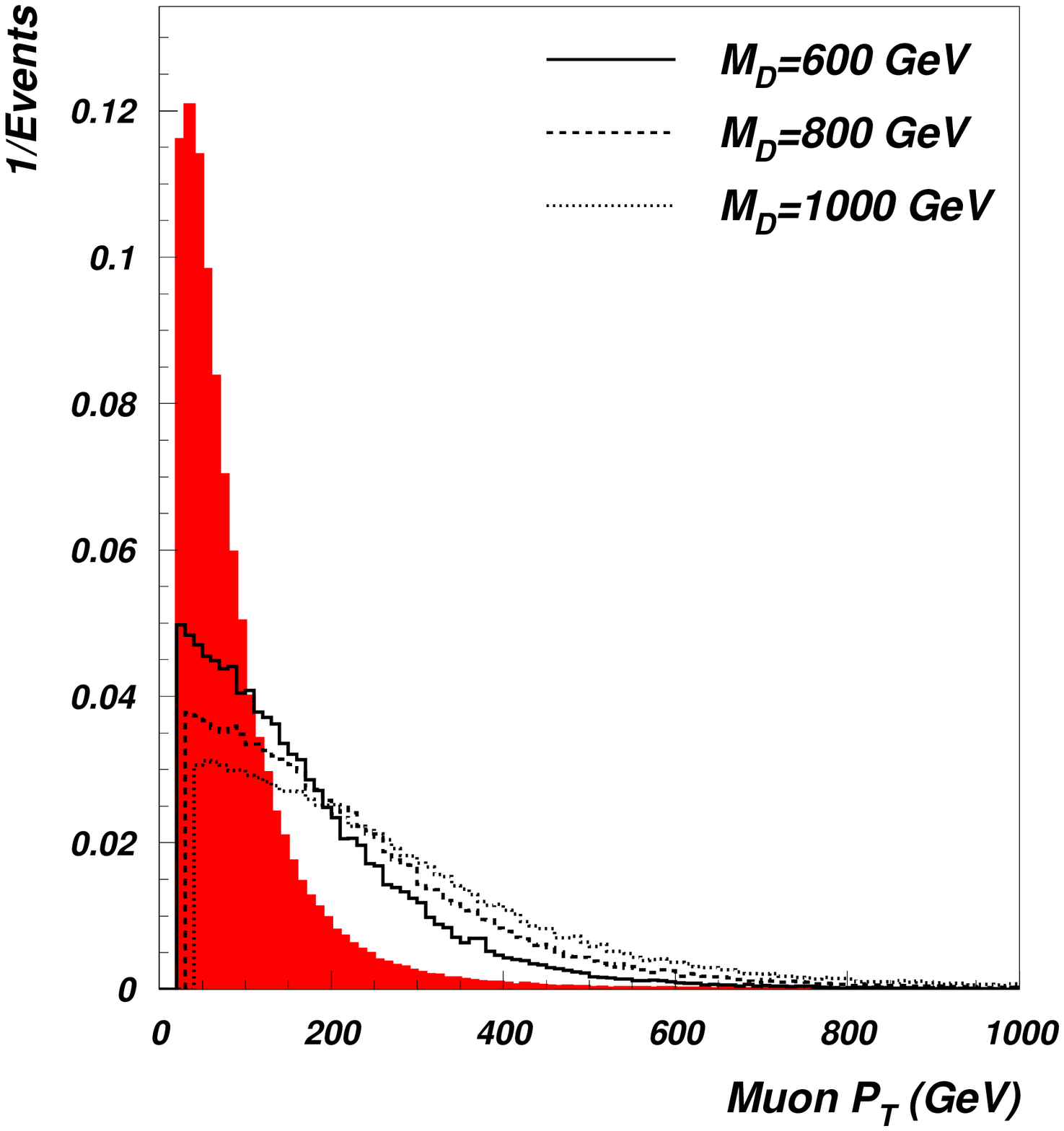}
\par\end{centering}

\caption{Jet (left), electron (center) and muon (right) transverse momentum
spectra for $3\ell+2j+E\!\!\!\!\!/_{T}$ channel. The shaded area
represents the SM background contribution.\label{fig:pt-CC1}}

\end{figure}

\begin{figure}
\begin{centering}
\includegraphics[scale=0.3]{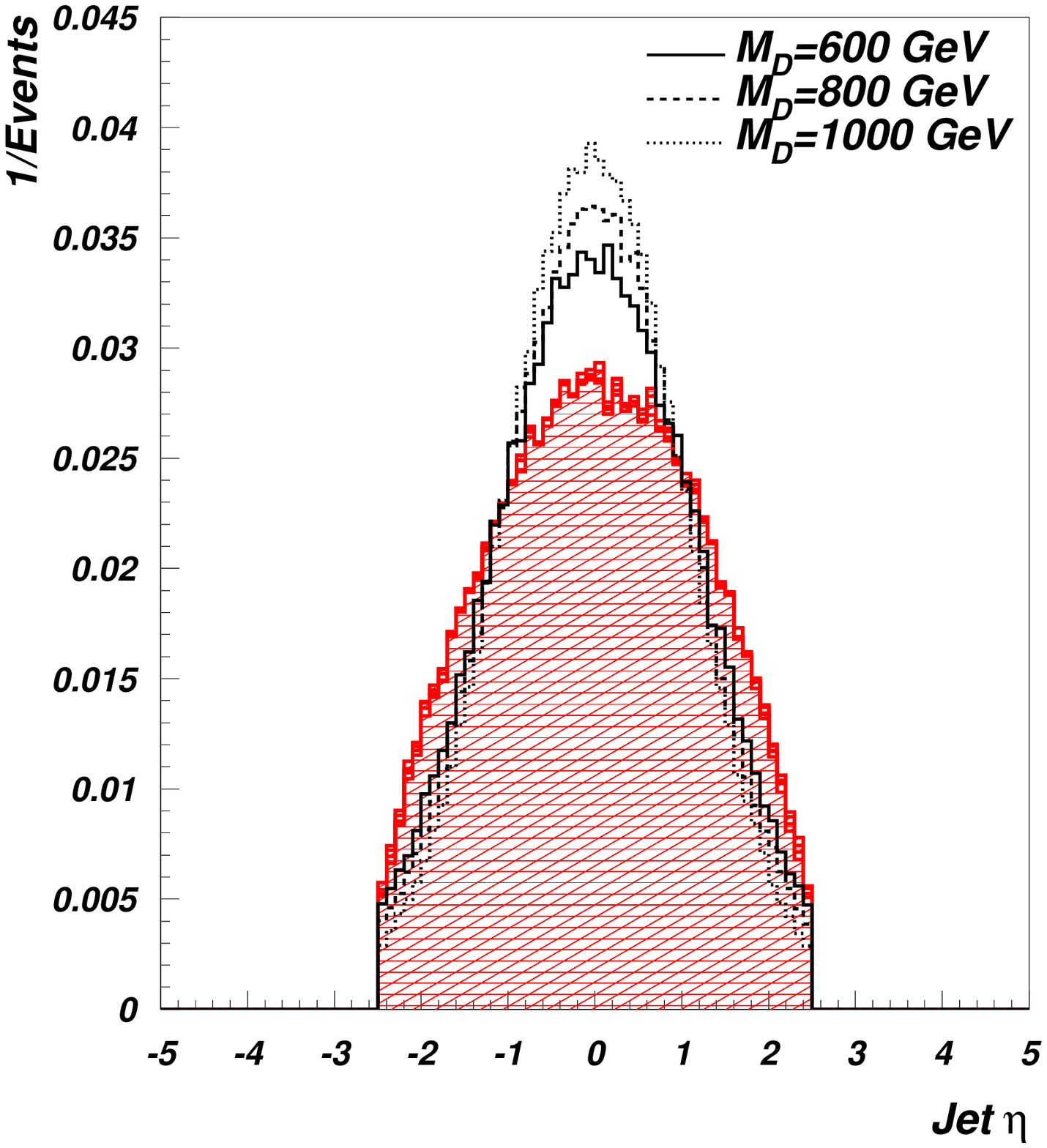}\includegraphics[scale=0.3]{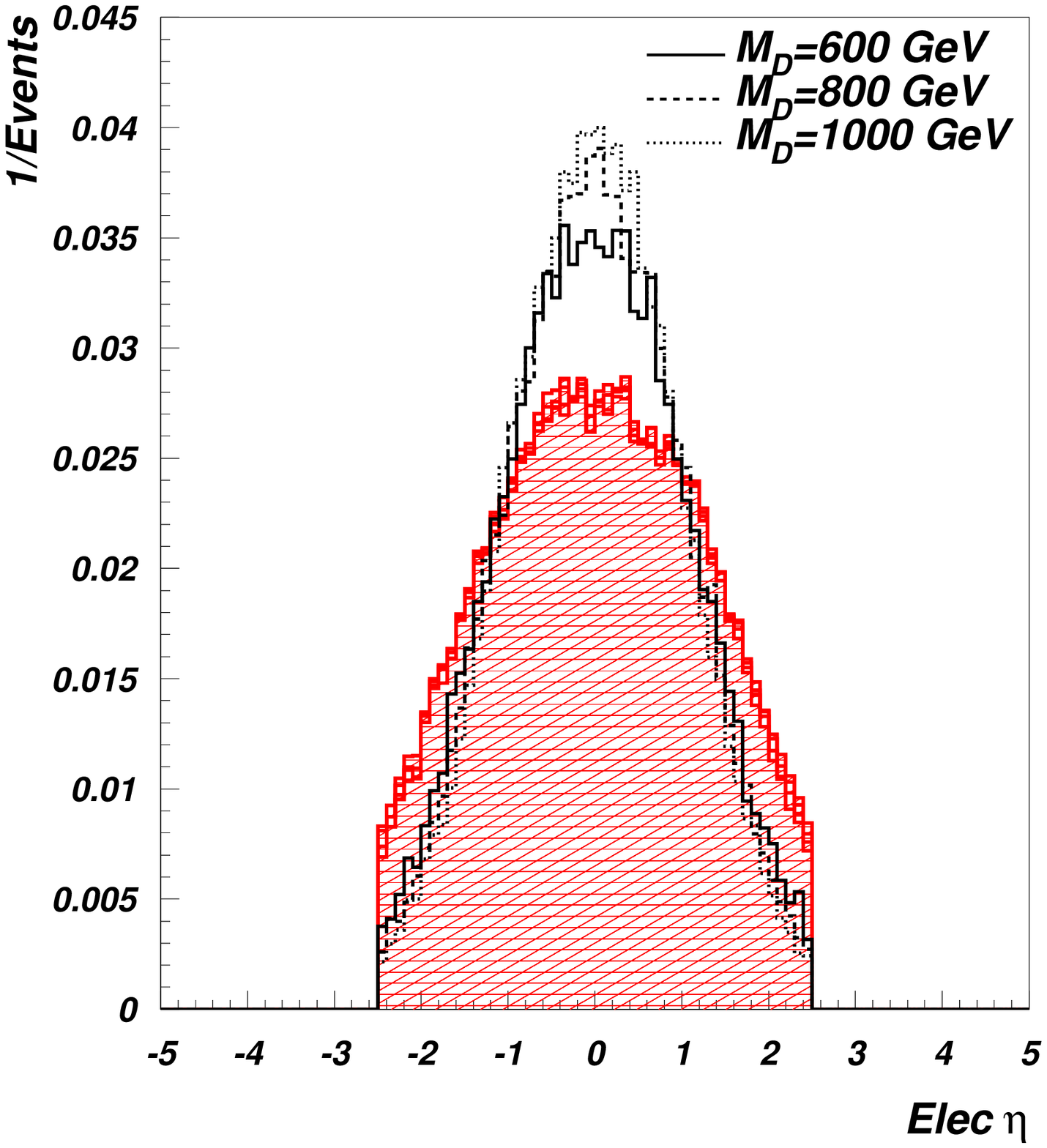}
\par\end{centering}

\caption{Jet (left) and electron (right) pseudorapidity distributions for different
$D$ quark masses using the $3\ell+2j+E\!\!\!/_{T}$ channel. The
shaded area represents the SM background contribution.\label{fig:eta-CC1}}

\end{figure}

\begin{figure}
\begin{centering}
\includegraphics[scale=0.3]{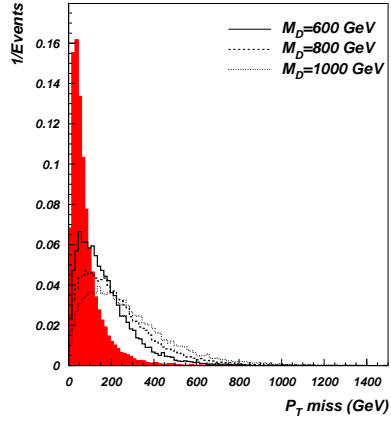}
\par\end{centering}

\caption{Missing $p_{T}$ distributions for different $D$ quark masses using
the $3\ell+2j+E\!\!\!\!\!/_{T}$ channel. The shaded area represents
the SM background contribution. \label{fig:missPt-CC1}}

\end{figure}

Transverse momentum spectra of jets, electrons and muons in the signal
events for various $D$ quark mass in comparison with the corresponding
spectra from the background events, are presented in Fig.$\;$\ref{fig:pt-CC1}
after the selection cuts defined above. One can see that the signal
events feature more energetic jets and leptons than the background
data. Pseudorapidity distributions of jets and electrons in the same
events are shown on Fig.~\ref{fig:eta-CC1}. Partons from signal
events are produced mostly in the central region, more so as the $D$
quark mass increases. Missing transverse momentum distribution, produced
in the event as a result of the semi-leptonic $W$ decay is shown
on the Fig.~\ref{fig:missPt-CC1}. As it can be seen, signal events
produce more energetic $W$ bosons, which results in larger missing
transverse momentum vector with respect to the SM $WZjj$ events. 

The reconstructed invariant mass of the two leptons was required to
be in the ($90\pm20$)~GeV mass window to ensure their origin from
the $Z$ boson. The $W$ boson {}``visible mass'' was reconstructed
from the measured lepton momenta and the $E\!\!\!\!\!/_{T}$ in the
transverse plane, assuming negligible neutrino longitudinal momentum.
A mass window constraint of ($80\pm20$)~GeV was applied to the reconstructed
$M_{\ell\,\nu}^{rec}$ to select events with a $W$ boson.\emph{ }

In case A, the identification of the two leptons originating from
the $Z$ boson decay forced the use of the third lepton to reconstruct
the \textbf{$W$} boson invariant mass. The method used to identify
these two leptons and thus to avoid the problems due combinatorics
related to the issue of lepton assignment to $Z$ or $W$, was to
select those giving the most accurate $Z$ boson mass. In case B,
as the leptons originating from $Z$ boson were of a different SM
family as compared to the one from the $W$ decay, reconstruction
of the $W$ invariant mass was straightforward. 

\begin{figure}
\begin{centering}
\includegraphics[scale=0.3]{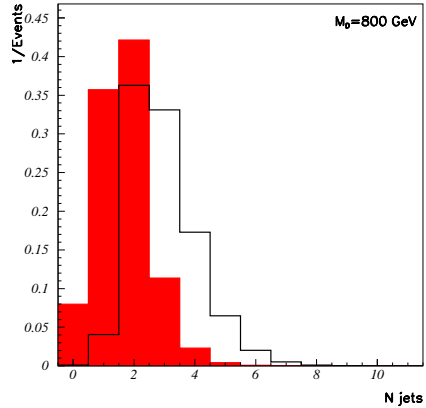}
\par\end{centering}

\caption{Number of jets with $p_{T}>70$ GeV for signal (white) and background
(shaded) events in the $3\ell+2j+E\!\!\!\!\!/_{T}$ channel for an
example $D$ quark mass of 800 GeV.\label{fig:Njets-CC1}}

\end{figure}

\begin{figure}
\begin{centering}
\includegraphics[bb=30bp 280bp 510bp 550bp,clip,scale=0.5]{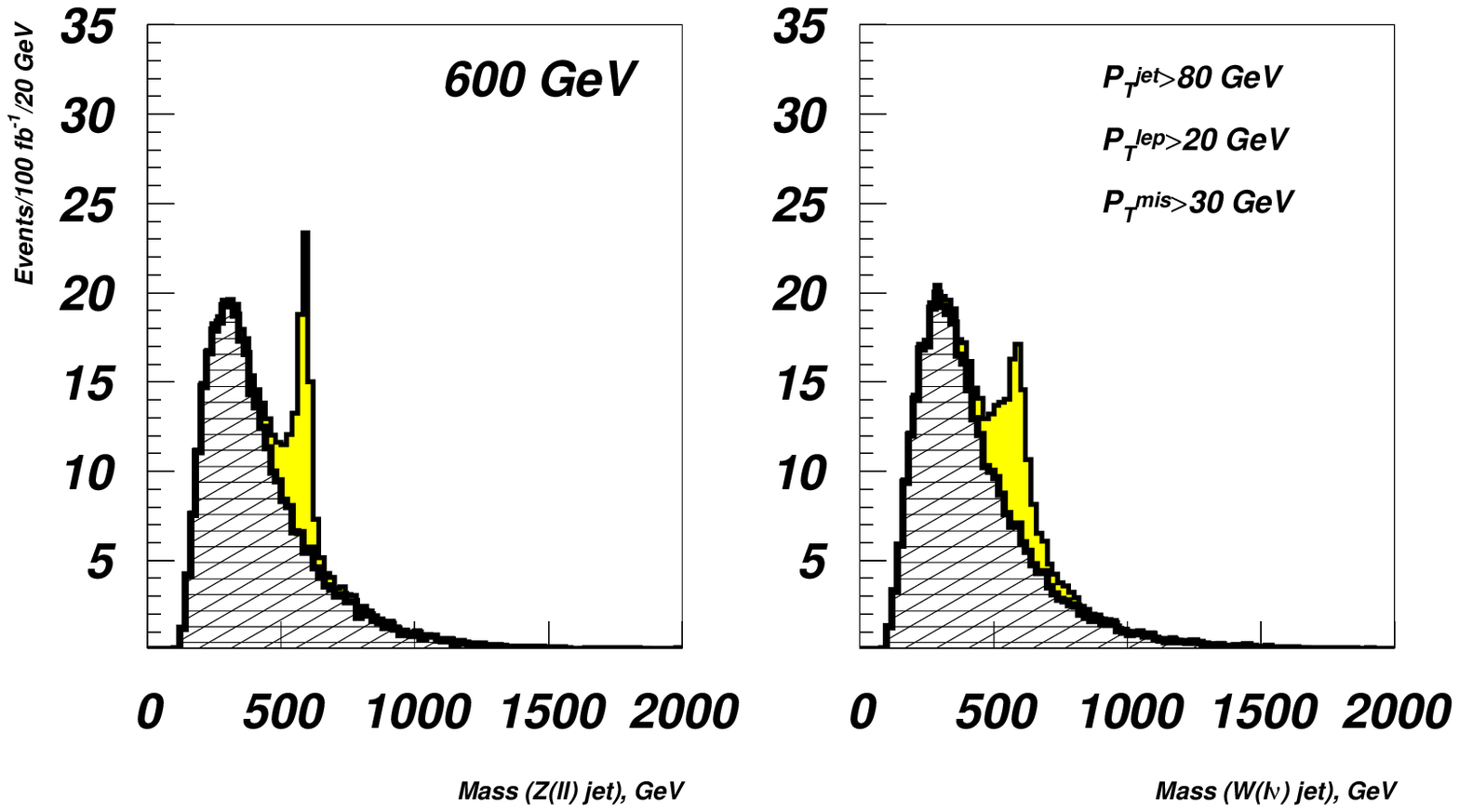}\includegraphics[bb=30bp 280bp 510bp 550bp,clip,scale=0.5]{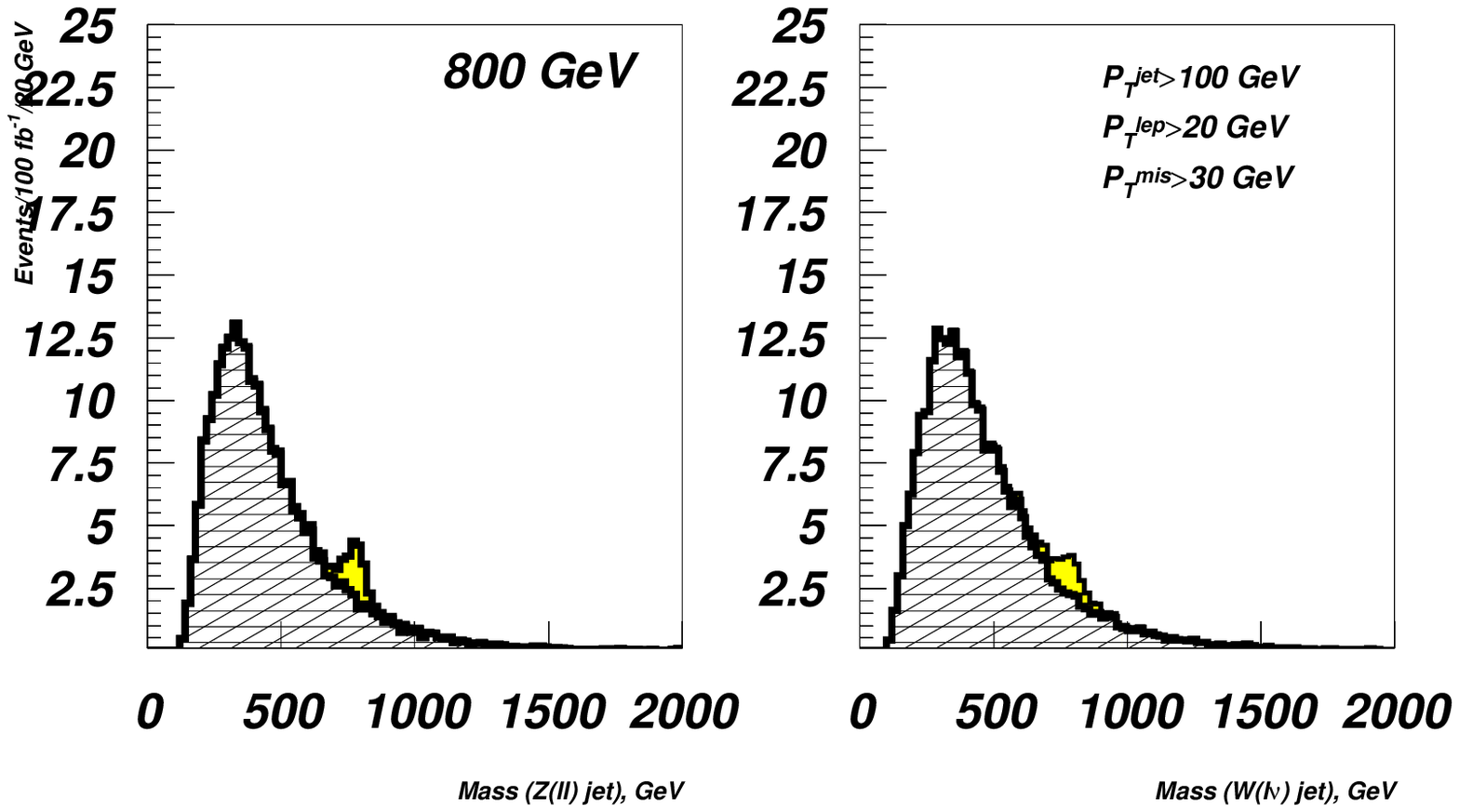}
\par\end{centering}

\caption{Invariant mass reconstruction spectra of signal (shaded) and background
(hashed) for two values of $D$ quark in $3\ell+2j+E\!\!\!\!\!/_{T}$
channel. The same lepton flavor was found in $Z$ and $W$ decay chains
(Case A).\label{fig:Signal-CC1-3lep}}

\end{figure}

\begin{figure}
\begin{centering}
\includegraphics[bb=30bp 280bp 510bp 550bp,clip,scale=0.5]{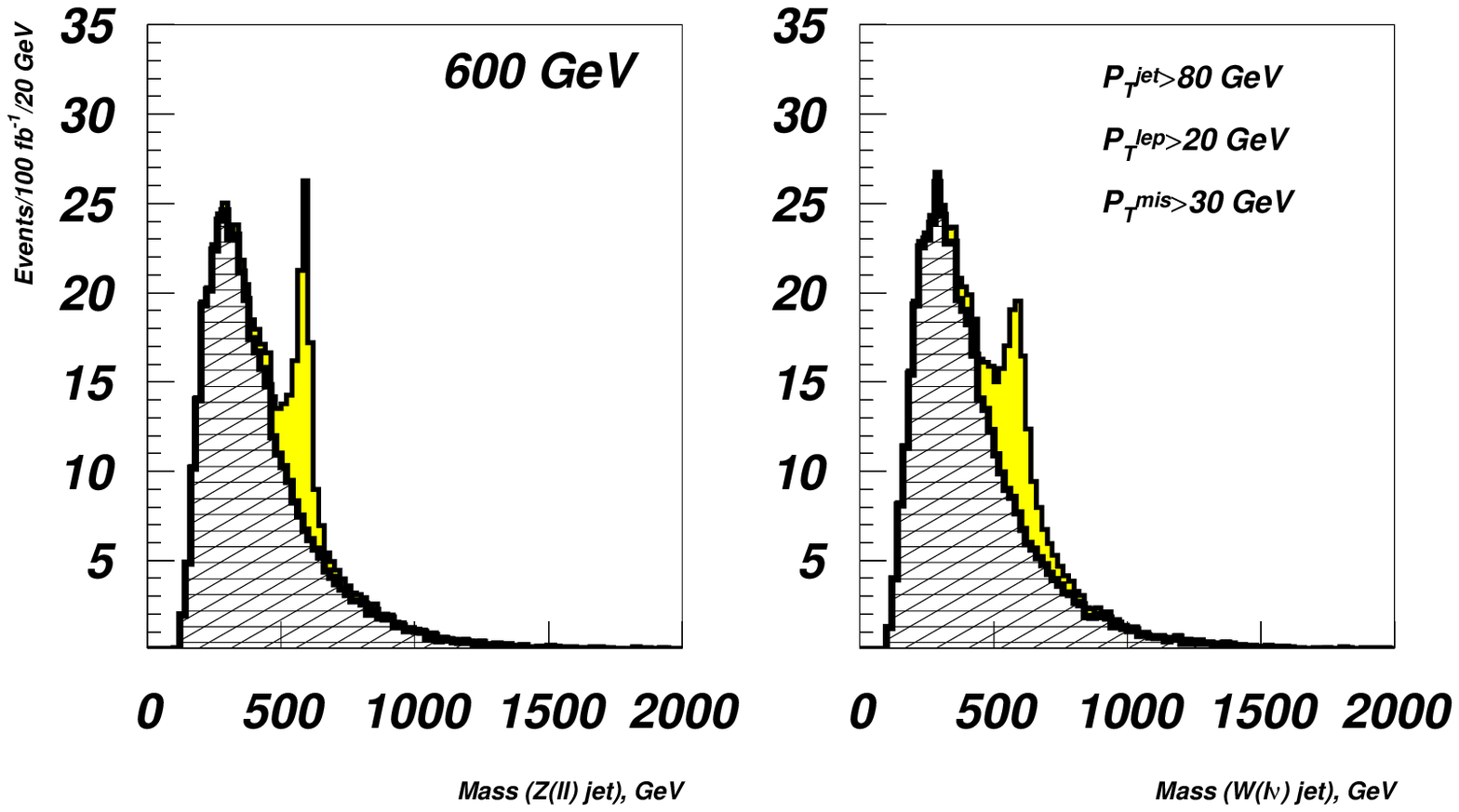}\includegraphics[bb=30bp 280bp 510bp 550bp,clip,scale=0.5]{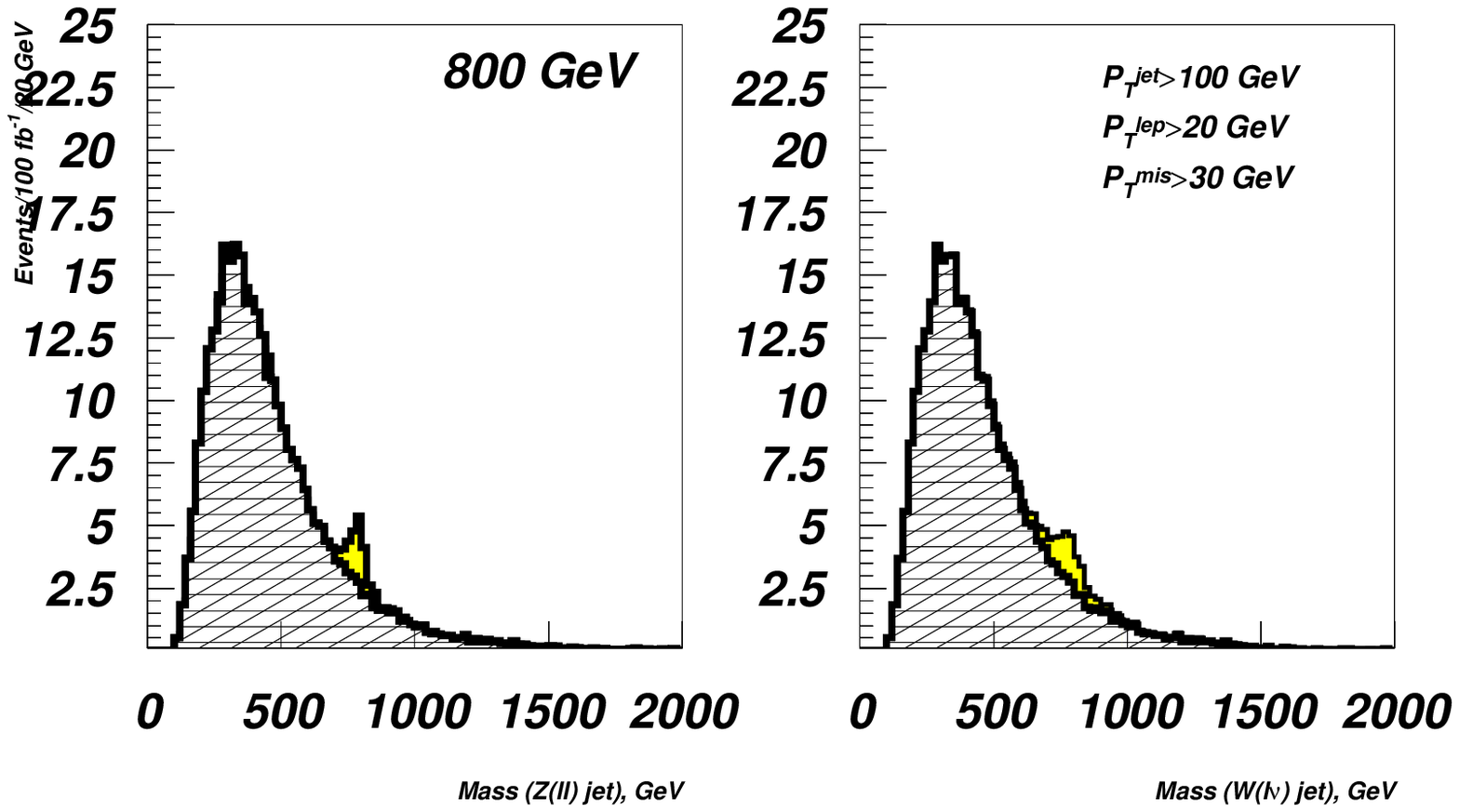}
\par\end{centering}

\begin{centering}
\includegraphics[bb=30bp 280bp 510bp 550bp,clip,scale=0.5]{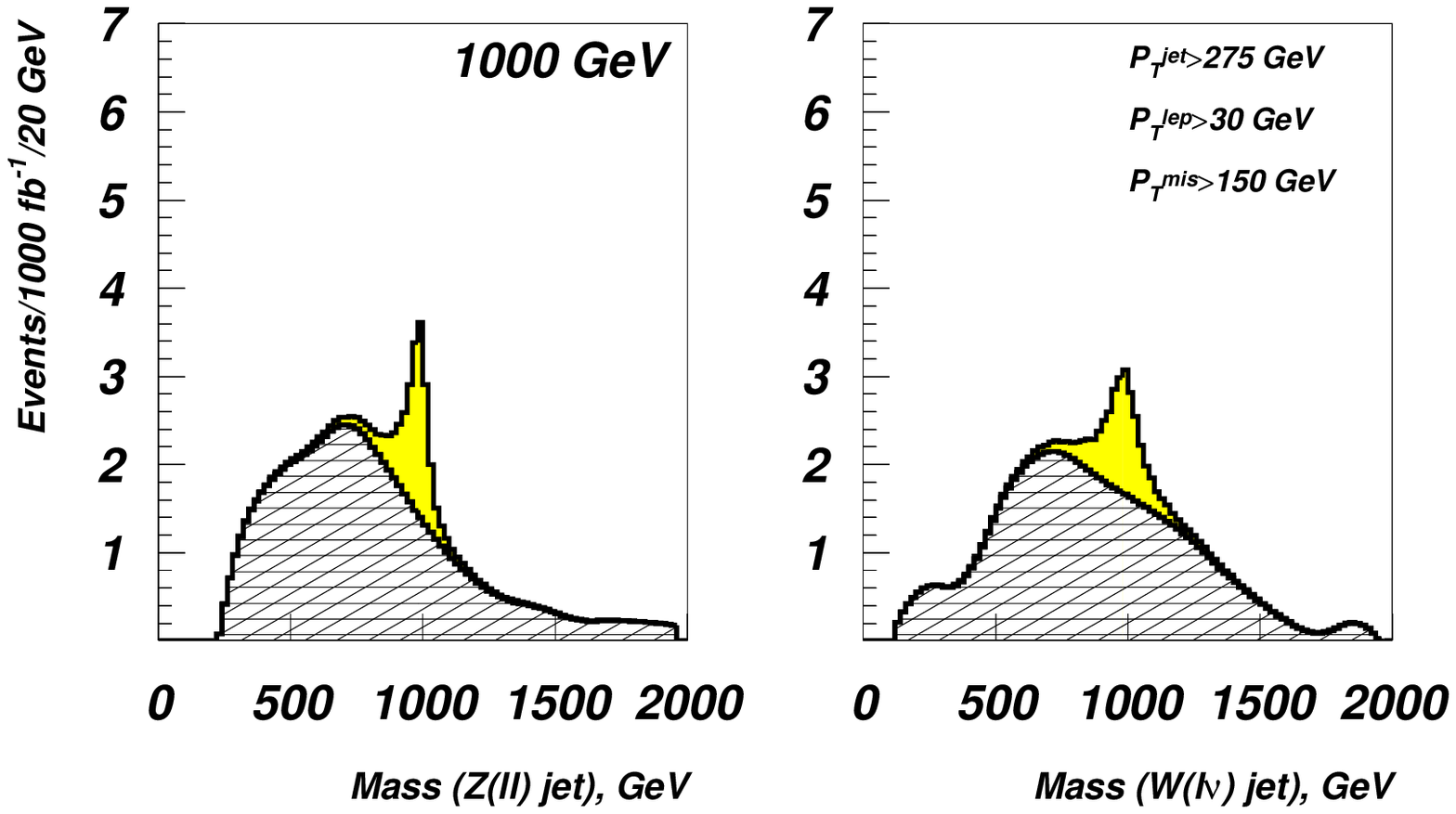}
\par\end{centering}

\caption{Invariant mass reconstruction spectra of signal (shaded) and background
(hashed) for three values of $D$ quark in $3\ell+2j+E\!\!\!\!\!/_{T}$
channel. The lepton flavor found in $Z$ and $W$ decay chains was
different (Case B). \label{fig:Reco-D-CC1}}

\end{figure}

As QCD radiation from energetic partons may result in an increased
number of produced jets, a correct assignment of selected jets might
not always be possible due to combinatorics. Moreover, the signal
events produce a significant number of jets, as can be seen from Fig.~\ref{fig:Njets-CC1},
where the jet ($p_{T}$> 70 GeV) multiplicity distributions for the
signal events of a $D$ quark mass 800~GeV and for the background
events are presented. Such a large jet multiplicity could pose a problem
for the identification of the jet originating from the $D$ quark
decay and hence for the reconstruction of its invariant mass. In both
cases the invariant mass of the $D$ quark, decaying to a $Z$ boson
and a jet was reconstructed by combining those two leptons with the
most energetic (or the next to most energetic) jets only. The ambiguity
in assigning the two most energetic jets to the two reconstructed
bosons can be solved by calculating the invariant mass of the $D$
quarks using all possible combinations and by taking the one which
gives the smallest mass difference between the $D$ quarks obtained
from $Z$ and $W$ in the same event. The $D$ quark invariant mass
reconstruction spectra obtained with this method are shown in Fig.\ref{fig:Signal-CC1-3lep}
for case A, where $Z$ and $W$ decay modes involve leptons of the
same generation and in Fig.\ref{fig:Reco-D-CC1} for case B, where
$Z$ and $W$ decay modes involve leptons of different generations.
The figures contain, for an integrated luminosity of 100~fb$^{-1}$
the reconstructed invariant mass distributions of $D\rightarrow Zj$
and $D\rightarrow Wj$ processes. The latter have wider invariant
mass spectra due to the use of the {}``visible mass'' of the $W$
in the reconstruction routine. 

\begin{figure}
\begin{centering}
\includegraphics[bb=10bp 10bp 510bp 567bp,scale=0.4]{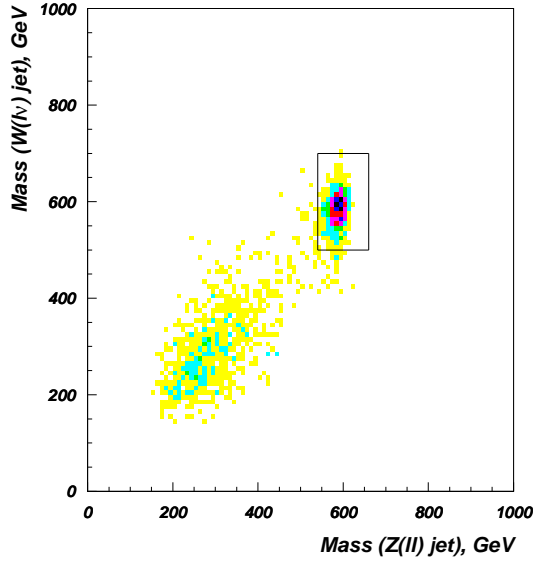}\caption{Example of the rectangular selection accepting events with two correctly
reconstructed $D$ quarks. The $Zj$ and $Wj$ acceptance widths for
different mass values are given in tables \ref{tab:signif-CC1-3lep}
and \ref{tab:signif-CC1}. Here the selection for type B events of
$m_{D}$=600 GeV is shown.}
\label{fig:rectangular-cuts}
\par\end{centering}
\end{figure}

The number of signal ($S$) and background ($B$) events together
with calculated statistical significance, corresponding to 100 fb$^{-1}$
integrated luminosity are presented in tables \ref{tab:signif-CC1-3lep}
and \ref{tab:signif-CC1} for type A and B events. The number of events
in the signal and background processes were calculated in sliding
mass windows of $\pm2\times\delta M$ around the peaks of the signal
events for both $Wj$ and $Zj$ invariant mass histograms. This selection
aimed to select the events where both $D$ quarks were successfully
reconstructed.\underbar{ }The example of the sliding windows mass
cuts to calculate the number of events in signal and the background
is presented in Fig. \ref{fig:rectangular-cuts} for type B events
and for a $D$ quark mass of 600 GeV. The selection windows for all
$D$ quark mass values are given in tables \ref{tab:signif-CC1-3lep}
and \ref{tab:signif-CC1} for type A and B events. The same tables
also contain the statistical significance of the signal calculated
as $S/\sqrt{B}$ if the total number of events were larger than 25.
Otherwise, the signal observation probability was calculated using
Poisson statistics and the significance of a normalized Gaussian distribution
yielding the same probability was reported.

\begin{table}
\caption{The expected number of events in mass windows of width $\delta M$
around the peak and the statistical significance for the $3\ell+2j+E\!\!\!\!\!/_{T}$
channel, after the cuts given in \ref{eq:cuts_3l_2j}, for type A
(same lepton generation) using 100 fb$^{-1}$ of integrated luminosity.\label{tab:signif-CC1-3lep}}

\begin{centering}
\begin{tabular}{c|c|c|c}
Mass (GeV) & 600 & 800 & 1000\tabularnewline
\hline
\hline 
$\delta$M ($D$$\rightarrow$ $Z$ jet)  & 120 GeV & 140 GeV & 160 GeV\tabularnewline
$\delta$M ($D$$\rightarrow$ $W$ jet) & 200 GeV & 220 GeV & 240 GeV\tabularnewline
\hline 
$S$ & 45 & 8.3 & 0.5\tabularnewline
$B$ & 11.4 & 4 & 0.25\tabularnewline
\hline 
significance & 13.3 & 3.3 & -\tabularnewline
\end{tabular}
\par\end{centering}
\end{table}

\begin{table*}
\caption{The expected number of events in the mass window around the peak and
the statistical significance for the $3\ell+2j+E\!\!\!\!\!/_{T}$
channel, after the cuts given in \ref{eq:cuts_3l_2j}, for type B
(different lepton generations) using 100fb$^{-1}$ of data.\label{tab:signif-CC1}}

\begin{centering}
\begin{tabular}{c|c|c|c}
Mass (GeV) & 600 & 800 & 1000\tabularnewline
\hline
\hline 
$\delta$M ($D$$\rightarrow$ $Z$ jet) & 120 GeV & 140 GeV & 160 GeV\tabularnewline
$\delta$M ($D$$\rightarrow$ $W$ jet) & 200 GeV & 220 GeV & 240 GeV\tabularnewline
\hline
$S$ & 52 & 10 & 1.3\tabularnewline
$B$ & 13.5 & 5 & 0.8\tabularnewline
\hline 
significance & 14.2 & 3.8 & -\tabularnewline
\end{tabular}
\par\end{centering}
\end{table*}

\subsection{Search using the $2\ell+4j$ channel}

This channel has one $D$ quark decaying through a $Z$ boson which
is expected to be reconstructed from two leptons (electrons and muons)
and a second $D$ quark decaying either via a $W$ boson or a second
$Z$ boson which can also be reconstructed using two energetic jets.
One should note that the reconstruction efficiency of a highly boosted
vector boson decaying hadronically can be lower compared to smaller
boosts since the small opening angle between the jets can lead to
a single jet. The signal events were simulated for both the $Z\, W\, j\, j$
and $Z\, Z\, j\, j$ cases. For this type of process, the signal is
sought in a final state made of a pair of energetic electrons (or
muons) in a multijet environment which includes two high $p_{T}$
jets. The background events consisted of the SM $WZ+2j$ , $ZZ+2j$
and $Z+2j$ processes, with generator level cross sections of 6.19,
5.5 and 25 pb respectively with the generator level cuts in Eq. (\ref{eq:generator level cuts})
except $p_{T,\, p}>50$ GeV.

\begin{figure}
\begin{centering}
\includegraphics[bb=10bp 20bp 510bp 567bp,clip,scale=0.3]{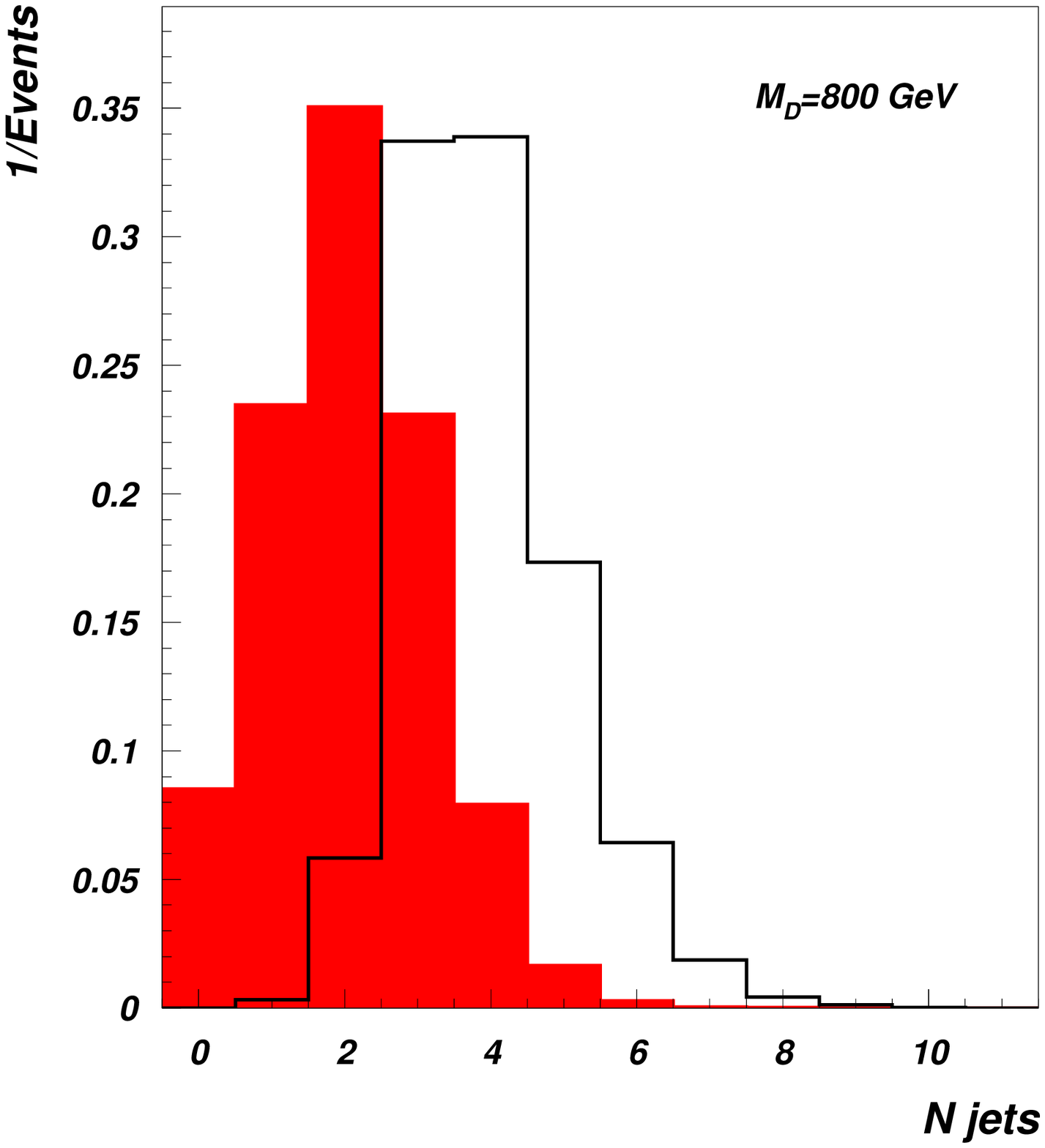}\includegraphics[bb=0bp 0bp 567bp 547bp,clip,scale=0.3]{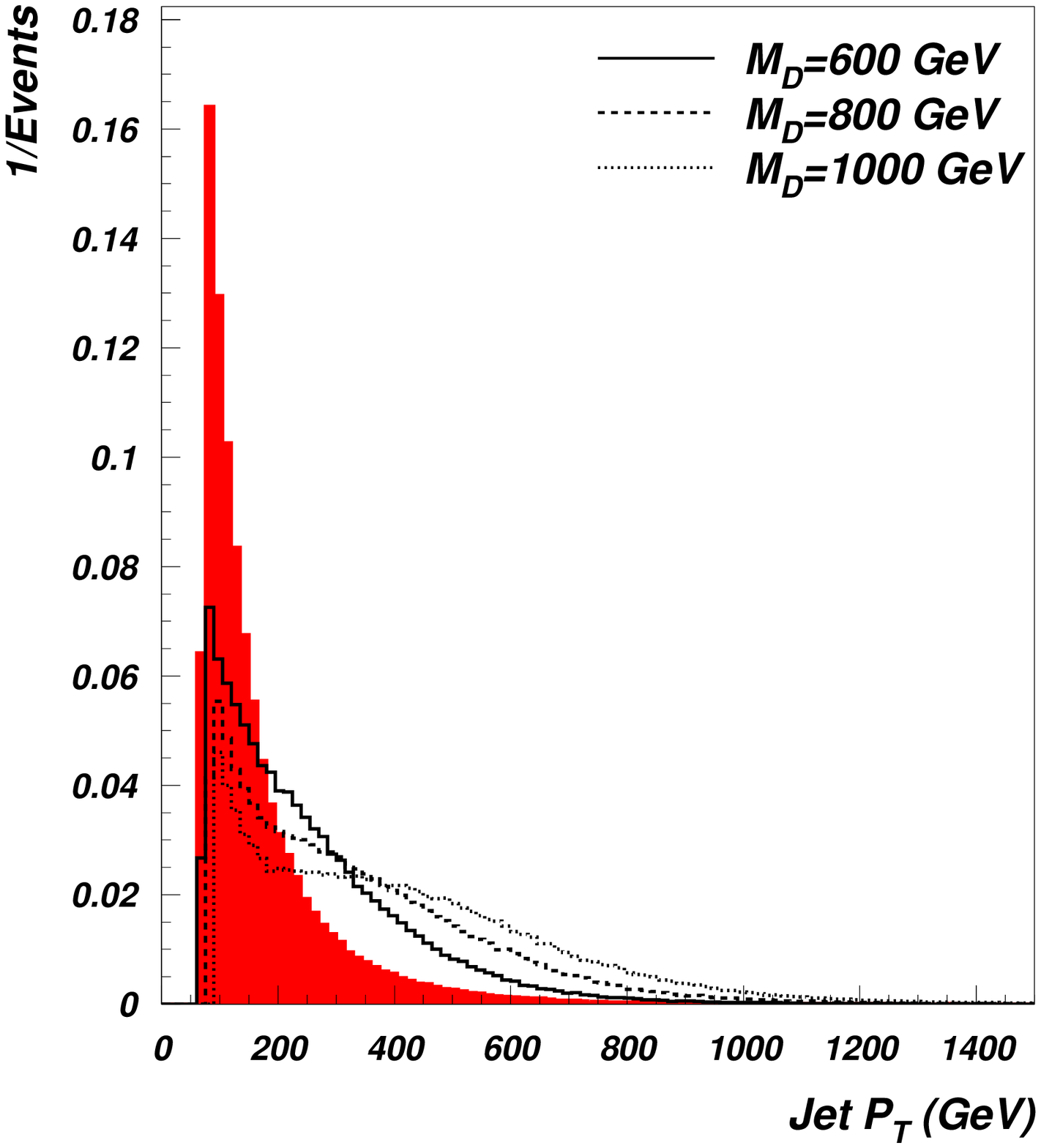}
\par\end{centering}

\caption{Multiplicity and $p_{T}$ jets for signal(white) and background(shaded)
events for $2\ell+4j$ channel \label{fig:Mult-Pt-Jet-CC2}}

\end{figure}

The transverse momentum distribution of the leptons in the final state
considered in this section do not differ much from those presented
in the previous section. Therefore, Fig. \ref{fig:Mult-Pt-Jet-CC2}
contains only the multiplicity of jets and their transverse momentum
spectra for signal events using various $D$ quark mass in comparison
with the corresponding spectra from background events. As it can be
seen in the same figure, the signal process produces more energetic
jets in the final state as compared to the SM background. However
the jet multiplicity is higher than the case considered in the previous
section. In general, high jet multiplicity creates problems in the
invariant mass reconstruction due to combinatorics.

The cuts for the reconstruction of the pair produced $D$ quarks with
mass of 600 GeV are presented in the set below:

\begin{eqnarray}
\#\ell(e,\,\mu) & = & 2\label{eq:2l_4j_cuts}\\
\#jet & \geq & 4\nonumber \\
p_{T,j} & > & 80\:\hbox{GeV}\nonumber \\
p_{T,\ell} & > & 20\:\hbox{GeV}\nonumber \\
|\eta(j,\,\ell)| & < & 2.5\nonumber \\
M_{\ell\ell}^{\hbox{rec}} & = & 90\pm20\:\hbox{GeV}\nonumber \\
M_{j\, j}^{\hbox{rec}} & = & 85\pm25\:\hbox{GeV}\nonumber \end{eqnarray}

For higher $D$ quark mass values of 800 and 1000 GeV, the jet transverse
momentum cuts were increased by 10 and 20 GeV respectively for each
next mass value to obtain a better value for signal significance ($S$/$\sqrt{B}$
or from Poisson statistics, depending on the available number of events
as mentioned before), wherever it was statistically possible. The
reconstruction of two lepton invariant mass was performed to obtain
the mass of $Z$ boson. 

\begin{figure}
\begin{centering}
\subfigure[$ZWjj$ with $m_D$=600 GeV]{\includegraphics[bb=20bp 270bp 510bp 555bp,clip,scale=0.5]{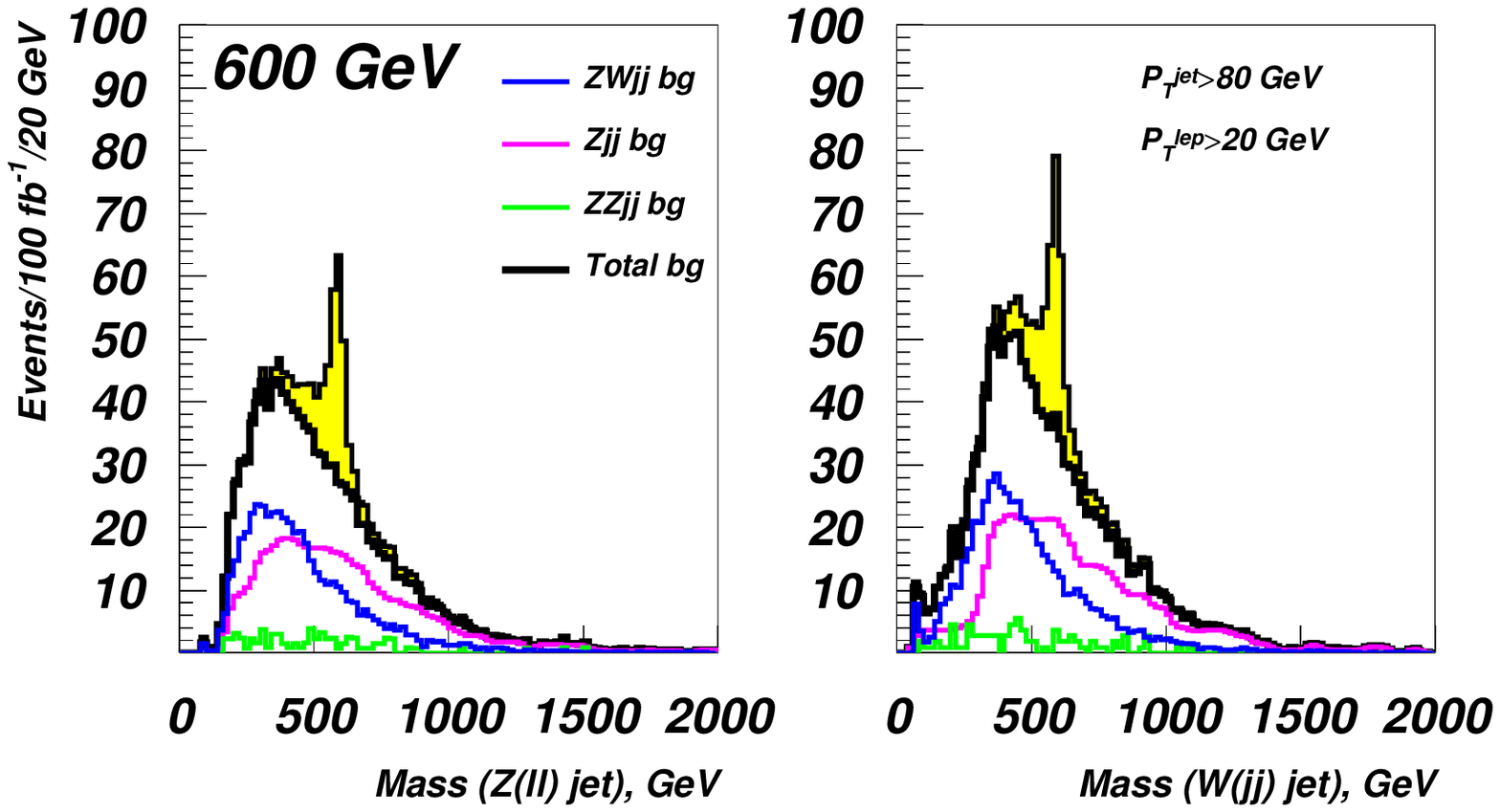}}\subfigure[$WZjj$ with  $m_D$=800 GeV]{\includegraphics[bb=20bp 270bp 510bp 555bp,clip,scale=0.5]{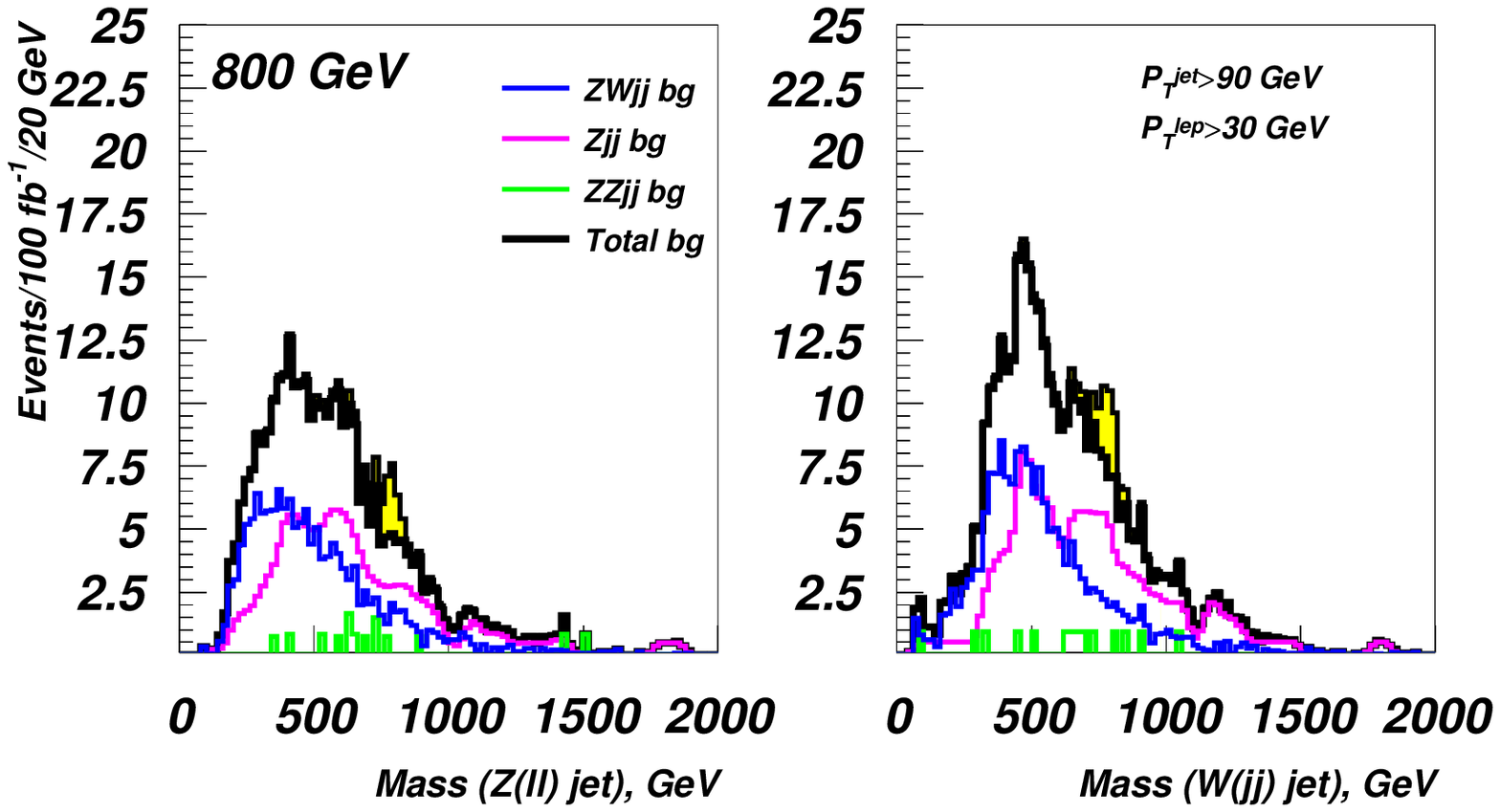}}
\par\end{centering}

\begin{centering}
\subfigure[$ZZjj$ with $m_D$=600 GeV]{\includegraphics[bb=20bp 270bp 510bp 555bp,clip,scale=0.5]{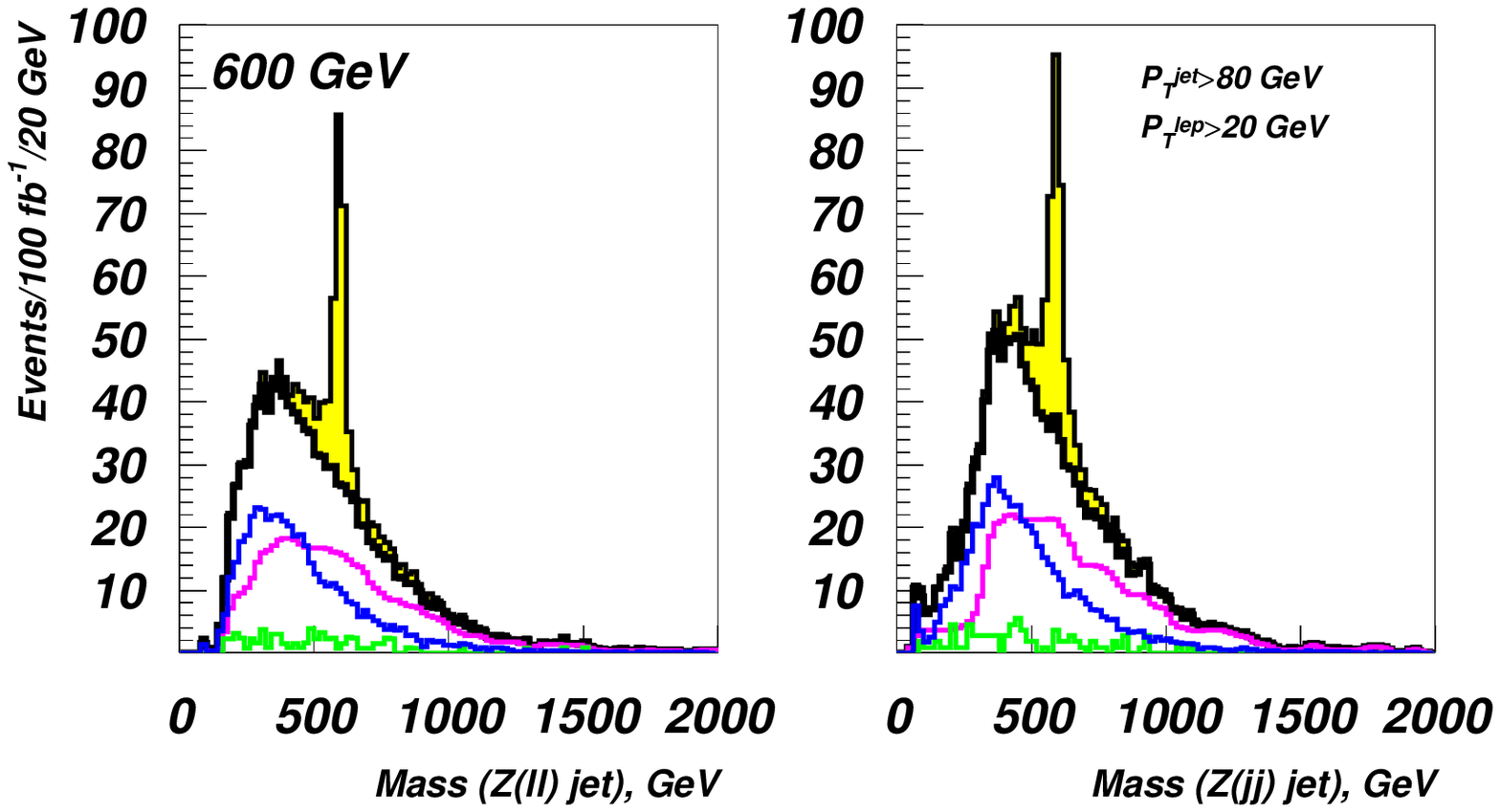}}\subfigure[$ZZjj$ with $m_D$=800 GeV]{\includegraphics[bb=20bp 270bp 510bp 555bp,clip,scale=0.5]{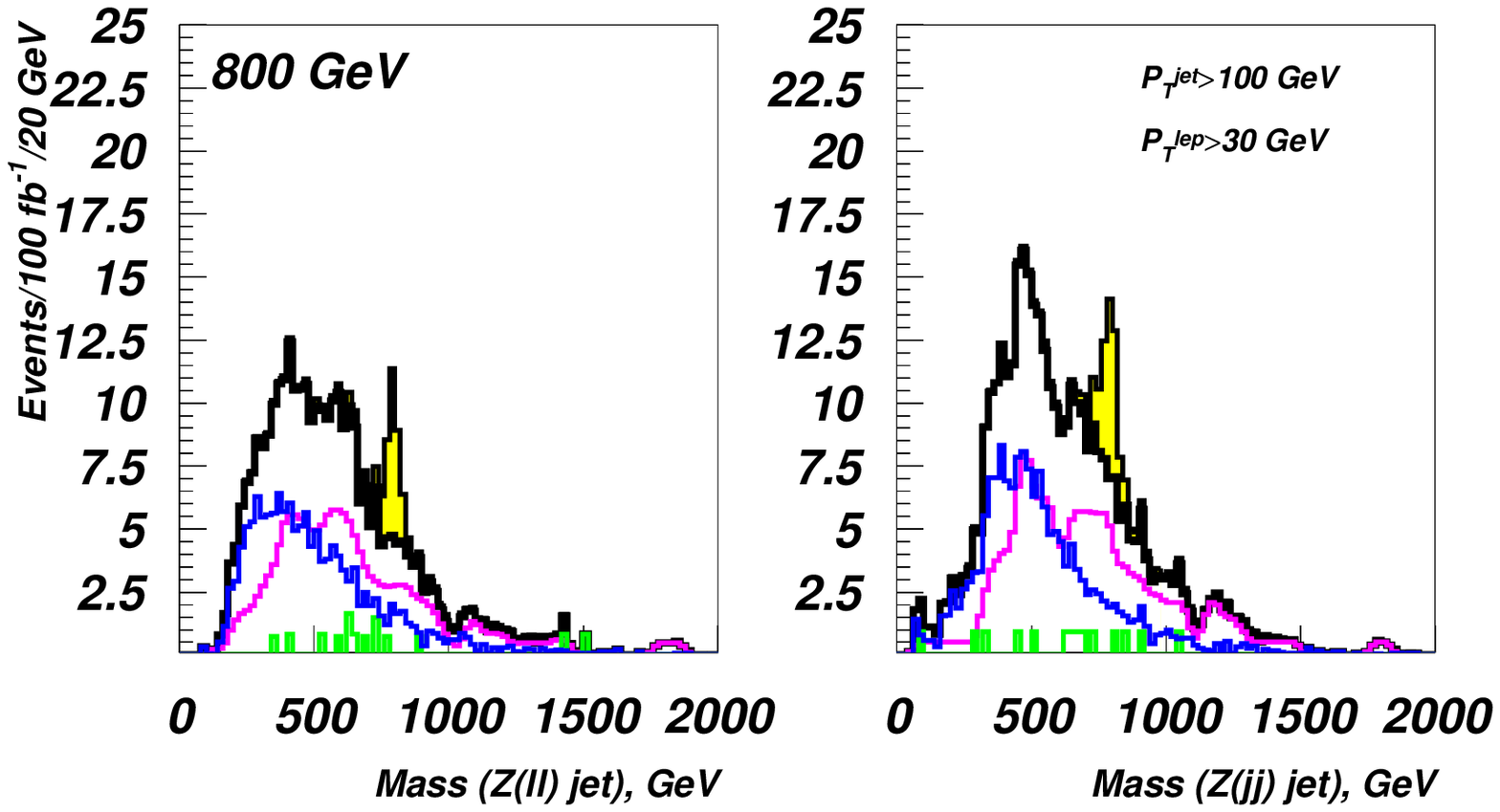}}
\par\end{centering}

\caption{Invariant mass reconstruction spectra of signal (shaded) and backgrounds
for $2\ell+4j$ channel\label{fig:Mass-reco-CC2}}

\end{figure}

The $D$ quark invariant mass reconstruction spectra with transverse
momentum cuts applied are shown in Fig. \ref{fig:Mass-reco-CC2}.
The mass of $D$ quark, decaying to a $Z$ boson and a jet was reconstructed
by combining two highest $p_{T}$ electrons (or muons) with a leading,
or a next to the leading $p_{T}$ jet choosing the case with the minimum
mass difference between the two $D$ quark candidates as in the previous
subsection. The invariant mass spectra for $D\rightarrow Z+jet$ decays
are shown on the left side of Fig. \ref{fig:Mass-reco-CC2} for each
considered mass. The decays of a $D$ quark to $W+jet$ were obtained
by requiring the reconstruction of $W$ mass with two jets in the
$(80\pm20)$~GeV window and then by combination of the two {}``$W$-tagged''
jets with the leading or  next to leading $p_{T}$ jet. The invariant
mass spectra of $D\to W+jet$ decays are presented on the right side
of the Fig. \ref{fig:Mass-reco-CC2} for each considered mass. The
distributions are shown for an integrated luminosity of 100 fb$^{-1}$.

The sum of number of signal and background events from both $W$ and
$Z$ involving cases with calculated statistical significance, corresponding
to one LHC year at high luminosity is presented in Table \ref{tab:signif-CC2}.
Numbers of events in the signal and background events were calculated
in the mass window of $\pm2\times\delta M$ around the respective
mass peak in the signal events. Since this decay channel has the biggest
branching ratio among all $D$ quark pair production processes, despite
the obvious challenge of multi-jet environment, the statistical significance
seems sufficient to observe $D$ quarks up to a mass of around 800
GeV.

\begin{table*}
\caption{The expected number of events in the mass window around the peak and
the statistical significance for the $2\ell+4j$ channel, after the
cuts given in \ref{eq:2l_4j_cuts}, for 100 fb$^{-1}$ of integrated
luminosity. \label{tab:signif-CC2}}

\begin{centering}
\begin{tabular}{c|c|c|c}
Mass (GeV) & 600 & 800 & 1000\tabularnewline
\hline
\hline 
$\delta$M ($D$$\rightarrow$ $Z$/$W$ jet) & 120 GeV & 140 GeV & 160 GeV\tabularnewline
\hline 
$S$ & 133 & 18 & 2.5\tabularnewline
$B$ & 9 & 3 & 2.8\tabularnewline
\hline
significance & 44.3 & 6.8 & 1.7\tabularnewline
\end{tabular}
\par\end{centering}
\end{table*}

\section{Results and Conclusions }

\begin{figure}[h]
\begin{centering}
\includegraphics[bb=20bp 25bp 555bp 545bp,clip,scale=0.4]{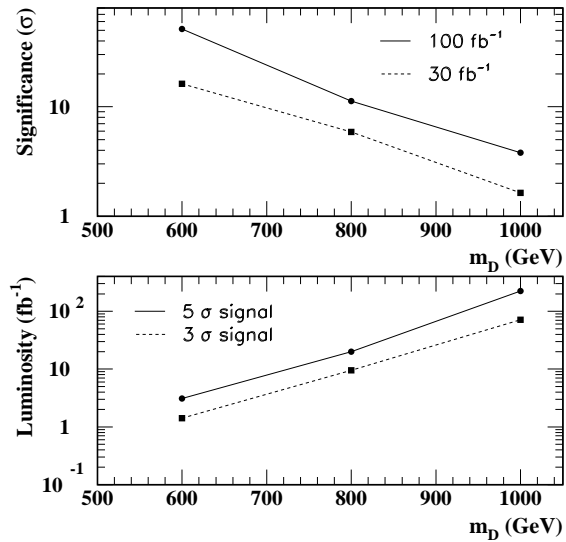}
\par\end{centering}

\caption{Upper plot: Combined significance as a function of the $D$ quark
mass for an integral luminosity of 30 (dashed line) and 100 (solid
line) fb$^{-1}$. Lower plot: Required luminosities for the 3$\sigma$
observation limit (dashed line) and the 5$\sigma$ discovery limit
as a function of the $D$ quark mass. \label{fig:final}}

\end{figure}

The analysis presented here has shown that the lightest of the quarks
predicted by the $E_{6}$ GUT models can be discovered in different
decay channels if its mass is less than 1 TeV. By combining multiple
channels from this study and from the previous one \cite{NC1-pub},
the discovery potential becomes higher and consequently the required
luminosity becomes lower. Table~\ref{tab:Yearly-signal-and} shows
the expected number of signal and background events from each channel
and the total significance after one year of running at nominal LHC
luminosity of 100 fb$^{-1}$ per year. To combine the significance
of the different channels, we have used a simple method. For each
channel, after obtaining the probability of compatibility with a background
hypothesis using Poisson or Gauss statistics depending on the available
number of events, we calculated the total probability for background
compatibility  of all four considered channels. The total probability
was converted to significance assuming a normalized Gauss distribution.
The top plot in Fig.~\ref{fig:final} shows the combined significance
for 30 and 100 fb$^{-1}$ of integrated luminosity as a function of
the $D$ quark mass. The same figure, on the bottom part, contains
the required luminosity in fb$^{-1}$ to observe 3$\sigma$ and 5$\sigma$
signals also as a function of the $D$ quark mass. The required luminosity
was calculated assuming Poisson statistics and the total number of
events were rounded down to the nearest integer. The presence of at
least one event was required for any given channel to be included
in the total significance calculation. 

\begin{table}
\caption{Signal and background events for different channels and different
masses of the $D$-quark for 100 fb$^{-1}$ integrated luminosity.
Combined total significance is also given as the last row. \label{tab:Yearly-signal-and}}

\begin{centering}
\begin{tabular}{r|c|c|c}
$m_{D}(GeV)$ & 600 & 800 & 1000\tabularnewline
\hline
$4\ell+2j\quad$ Signal & 16 & 3.7 & 0.74\tabularnewline
background & 3.0 & 1.3 & 0.4\tabularnewline
-$\ln$ p & 21.47 & 4.78 & 1.44\tabularnewline
\hline 
$2\ell+2j+E\!\!\!\!\!/_{T}\quad$ Signal & 53 & 19 & 4\tabularnewline
background & 12 & 13 & 5\tabularnewline
-$\ln$ p & 120 & 15.81 & 3.32\tabularnewline
\hline 
$3\ell+2j+E\!\!\!\!\!/_{T}\quad$ signal & 97 & 18.3 & 1.8\tabularnewline
background & 24.9 & 9.0 & 1.05\tabularnewline
-$\ln$ p & 191.4 & 20.66 & 1.69\tabularnewline
\hline 
$2\ell+4j\quad$ signal & 133 & 18 & 2.5\tabularnewline
background & 9 & 3 & 2.8\tabularnewline
-$\ln$ p & 983 & 25.3 & 2.44\tabularnewline
\hline 
$-\Sigma$$\ln$ p & 1315.9 & 66.5 & 8.9\tabularnewline
Combined Significance ($\sigma$) & 51.3 & 11.3 & 3.8\tabularnewline
\end{tabular}
\par\end{centering}
\end{table}

In conclusion, we show that the combination of the studied channels
allows the discovery of the lightest isosinglet quark predicted by
the $E_{6}$ GUT model within one year of LHC running time if its
mass is up to about 950 GeV with a significance of 5$\sigma$ or more.
As pair production was considered, these results are independent of
the mixing angle between the $D$ quark and its SM counterparts provided
that the lightest iso-singlet quark mixes mainly to the first or second
SM families. Although this study is based on a fast simulation of
the detector response which was not fully validated and there are
uncertainties associated with the QCD scale, statistical errors etc,
we believe that the conservative selection cuts and the simplicity
of the reconstruction algorithms give reliability to the conclusions.
Additionally, the verification of these results using the full simulation
of the detector is underway as preparation to first data from the
LHC \cite{CSC}.

\subsection*{Acknowledgments}

The authors would like to thank L. Tremblet and CERN Micro Club for
kindly providing computational facilities, F. Ledroit, G. Azuelos
and G. Brooijmans of ATLAS experiment for useful discussions. R.M.
would like to thank A. Belyaev for his assistance in model calculations.
R.M. also thanks NSERC/Canada for their support. A.S.'s work has been
supported by a Marie Curie Early Stage Research Training Fellowship
of the European Community's Sixth Framework Programme under contract
number (MEST-CT-2005-020238). S.S acknowledges the support from the
Turkish State Planning Committee under the contract DPT2002K-120250.
G.U.'s work is supported in part by U.S. Department of Energy Grant
DE FG0291ER40679. This work has been performed within the ATLAS Collaboration
with the help of the simulation framework and tools which are the
results of the collaboration-wide efforts.

\end{document}